\documentclass{jfm}
\usepackage{graphicx}
\usepackage{epstopdf, epsfig}
\usepackage[dvipsnames]{xcolor}
\usepackage{amsmath}
\usepackage{gensymb}
\usepackage{subfigure}
\usepackage[normalem]{ulem}

\shorttitle{Visualizing the turbulent energy cascade }

\shortauthor{R. McKeown, A. Pumir, M. Brenner, S. Rubinstein, and R. Ostilla-M\'{o}nico}

\title{Energy transfer and vortex structures: Visualizing the incompressible turbulent energy cascade}

\author{Ryan McKeown\aff{1},
  Alain Pumir\aff{2},
  Shmuel M. Rubinstein\aff{3},
  Michael P. Brenner\aff{1},
 \and  Rodolfo Ostilla-M\'{o}nico\aff{4,5}
  \corresp{\email{rodolfo.ostilla@uca.es}}}

\affiliation{\aff{1}School of Engineering and Applied Sciences, Harvard University, Cambridge, MA 02138,  USA 
\aff{2}Univ Lyon, ENS de Lyon, Laboratoire de Physique, F-69342 Lyon, France
\aff{3}Racah Institute of Physics, The Hebrew University of Jerusalem, Jerusalem, Israel
\aff{4}Department of Mechanical Engineering, University of Houston, Houston, TX 77204, USA
\aff{5}Escuela Superior de Ingener\'ia, Universidad de C\'adiz, C\'adiz, Spain}

\begin{document}

\maketitle

\begin{abstract}
The transfer of kinetic energy from large to small scales is a hallmark of turbulent flows. Yet, a precise mechanistic description of this transfer, which is expected to occur via an energy cascade, is still missing.  Several conceptually simple configurations with vortex tubes have been proposed as a testing ground to understand the energy cascade. Here, we focus on incompressible flows and compare the energy transfer occurring in a statistically steady homogeneous isotropic turbulent (HIT) flow with the generation of fine-scale motions in configurations involving vortex tubes.  We start by filtering the velocity field in bands of wavenumbers distributed logarithmically, which allows us to study energy transfer in Fourier space and also visualize the energy cascade in real space. In the case of a statistically steady HIT flow at a moderate Reynolds number, our numerical results do not reveal any significant correlation between regions of intense energy transfers and vorticity or strain, filtered in corresponding wavenumber bands, nor any simple self-similar process. In comparison, in the transient turbulent flow obtained from the interaction between two antiparallel vortex tubes, we observe a qualitatively simpler organization of the intense structures, as well as of the energy transfer.
The process leading to vortex reconnection via flattening of interacting vortex cores in two intense ribbons of vorticity, also appears as qualitatively different from the physics of HIT. Our results indicate that the specific properties of the transient flows affect the way energy is transferred, and may not be representative of HIT.
\end{abstract}


\section{Introduction}
The cornerstone of our understanding of turbulence rests on the
notion that for three-dimensional incompressible flows at a very high Reynolds number, energy is transferred from the large scale at which the flow is stirred down to very small scales where fluid motion is damped by   viscosity~\citep{Taylor35,vKH38,K41}.
\cite{Richardson:1922} famously described the cascade in terms of whirls that generates ever smaller whirls, down to a scale where viscosity prevents any further small-scale generation. This underlying self-similar picture has since become widely accepted, at least as a theoretical starting point \citep{Frisch:1995}. A precise description of the mechanisms leading to the cascade, however, has remained elusive.  

The importance of understanding
how energy cascades down to the dissipative scale cannot be overstated. The question arises naturally in order to model the decay of turbulence~\citep{Batchelor:1947,Batchelor:1948a,Batchelor:1948b,Moffatt:2002}, and, by extension, the decay of turbulence generated by a grid in a wind tunnel~\citep{Ueberoi:1963,vAtta:1968,Warhaft:1978,Sreenivasan:1980,Antonia:2003,Sinhuber:2015}. Additionally, the development of subgrid-scale models
in the context of large-eddy simulations (LES), one of the most used numerical techniques in fluid dynamics, amounts to providing a parameterization of the energy transfer from the resolved to the unresolved scales of a flow in a numerical simulation~\citep{Smagorinsky:1963,Leonard:1975,Bardina:1980,Germano:1991,Meneveau:1994,Metais:1996}.
Due to its significance, this problem has been extensively studied, even in the simplest case of homogeneous, isotropic turbulence, both experimentally~\citep{Liu:1994,Tao:2002,Katz:2010} and numerically~\citep{Menon:1994,Kerr:1996,Meneveau:2000}.
One of the main research questions is quantifying the locality of the energy transfer in Fourier space~\citep{Kraichnan:1966,Zhou:1993,Mininni:2005,Domaradzki:2009,Aluie:2009,Eyink:2009}. Detailed studies of the contribution of triads of wavenumbers in the energy transfer point to a subtle structure: some of the triads contribute to a nonlocal energy transfer~\citep{Domaradzki:1990}, with possible consequences to the structure and dynamics of the smallest eddies in the flow~\citep{Brasseur:1994,Yeung:1995}. The locality of the energy transfer in Fourier space can also be directly studied by representing the flux of energy between wavenumber bands~\citep{Favier:2014,Verma:2018,Alexakis:2005}, a method which we will use extensively here. We note that the analysis in terms of wavenumber bands can also be used for more complex flows, e.g. involving convection and rotation~\citep{Favier:2014,Kunnen:2016}, and in the case of magnetohydrodynamic turbulence, this approach reveals subtle exchanges between kinetic and magnetic energy~\citep{Mininni:2005,Alexakis:2005,Aluie:2010,Teaca:2011}. 

Aside from these applications, the notion of energy transfer is also essential to describe how an ordered, laminar flow evolves in time to become turbulent, by generating motion at increasingly small scales.
In this context, visualization tools, from experiments as well as from direct numerical simulations (DNS) of the Navier-Stokes equations, have proven to be an invaluable tool to identify the physical mechanisms governing the formation of small-scale flow structures.
It was recently realized that the elementary configuration of two colliding vortex rings~\citep{Lim:1992,McKeown:2018,McKeown:2020} leads, at sufficiently high Reynolds numbers, to the development of a disordered flow with a $k^{-5/3}$ energy spectrum, consistent with Kolmogorov prediction for turbulence~\citep{K41}. Remarkably, this spectrum results from a cascade of instabilities~\citep{Tao:2016,Brenner:2016} during the interaction between the vortices, which is responsible for the formation of small-scale flow structures~\citep{McKeown:2020}.
The instability that mediates the cascade, known as the elliptical instability~\citep{Kerswell:2002,Leweke:2016}, is induced by the deformation of the core of each filament due to the strain generated by the opposing vortex~\citep{McKeown:2020}. 
In another related flow, i.e.~the reconnection between two vortex filaments at sufficiently high Reynolds numbers, \cite{Yao:2020} observed the development of small-scale fluid motion with a Kolmogorov $k^{-5/3}$ spectrum and also suggested an ``avalanche'' of iterative reconnections as a possible mechanism for the cascade.

An understanding of the underlying energy transfer mechanisms in these conceptually simple flow configurations could possibly lead to insights on the cascading motion in turbulence~\citep{Ostilla:2021,Yao:2022}. 
In fact, the complex motion in turbulent flows has been conjectured to involve elementary coherent structures~\citep{Hussain:1986}; vortex tubes are likely candidates to capture these important dynamical processes~\citep{Siggia:1981,Jimenez:1993,Goto:2008,Buaria:2020}. This has provided an essential motivation for the study of the interactions of vortex tubes~\citep{Siggia:1985,Brenner:2016,Moffatt:2019a} and suggests the use of a common approach to study the transfer of energy in both fully turbulent flows as well as various configurations of interacting vortex tubes. 

Comparing energy transfer between turbulent flows and these various configurations is one of the main objectives of this work.
As already mentioned, one of the assumptions often used is that the transfer of energy occurs in a self-similar manner. This hypothesis is clearly made in the phenomenological approach of~\cite{Tennekes:1972}, who derived from elementary fluid mechanical considerations an explicit scaling expression for the energy transferred between scales. Consistent with the findings of~\cite{McKeown:2018,McKeown:2020}, the work of \cite{Yoneda:2021} also postulates a hierarchical structure of vortices to describe vortex stretching, inspired by the numerical results of \cite{Goto:2017} and \cite{Motoori:2019}. 

To investigate these issues, we proceed by decomposing the velocity field into logarithmically spaced shells of wavenumbers, $I_P$, 
such that $ 2^{P-1} k_f/\sqrt{2} \le | \mathbf{k} | \le 2^{P-1} k_f \times \sqrt{2}$, where $k_f$ is a fixed wavenumber corresponding to the largest scale flow structure. 
The use of logarithmic bands of wavenumbers, rather than linear ones (as in \cite{Zhou:1993,Alexakis:2005} for example) was chosen as it is particularly appropriate to study a self-similar problem, as turbulence has often been postulated to be. 
We introduce the corresponding band-passed filtered velocity fields $\mathbf{u}_P$, from which we can define associated quantities such as the vorticity, rate of strain, and kinetic energy.
In particular, we will focus on the rate of energy transferred from the band of modes $I_K$ to the band $I_Q$ as $\mathcal{T}_{K,Q} = -\mathbf{u}_Q \cdot (\mathbf{u} \cdot \nabla ) \mathbf{u}_K$ (c.f.~below and \cite{Mininni:2006} for the full derivation of this quantity).
This equation allows us to identify the regions of the flow where $\mathcal{T}_{K,Q}$ is positive, indicating that energy is transferred from the band $I_K$ to $I_Q$. We can then use this to correlate the regions in real space where the flow transfers energy with the presence of regions of high vorticity or high strain in the flow and elucidate the mechanisms for energy transfer across scales.
In the elementary case of homogeneous isotropic turbulent (HIT) flows, this method allows us to rule out any significant correlation between energy transfer and regions of intense vorticity or strain in HIT, a situation that was not substantially changed by imposing the large-scale structure of the flow suggested by~\cite{Goto:2017}. In contrast, we find much stronger correlations between transfer and vorticity and strain in the case of turbulent generated by the interaction of two antiparallel vortices. 

Our work is organized as follows. In the Methods section (Sec.~\ref{sub:methods}), we present our approach to studying energy transfer and describe our numerical methods.  The case of a statistically stationary homogeneous turbulent flow at moderately large Reynolds number ($Re_\lambda = 210$) is considered in Section~\ref{sec:HIT}. In particular, we recover the locality properties of the energy transfer in Fourier space~\citep{Kraichnan:1966,Tennekes:1972,Zhou:1993,Aluie:2009}. In  Section~\ref{sec:t-dependent-Fourier}, we consider the case of statistically unsteady flows, starting from the simple configuration of two initially antiparallel vortex tubes.
We analyze the generation of small-scales in the flow in Fourier space and show that intense energy transfer correlates with the locations of filaments with intense vorticity. This correlation persists, up to the point where the flow develops a turbulent regime, with a $k^{-5/3}$ spectrum. We demonstrate, however, that this simple correlation between the observed vortical structures and energy transfer scales does not hold when imposing a large-scale flow structure, consisting of four alternating vortex tubes in Section~\ref{sec:Goto}. Finally, we
contextualize our results with a recent analysis of vortex stretching, and we present our conclusions in Section~\ref{sec:concl}.
 
\section{Methods}
\label{sub:methods}

\subsection{Analysis of energy transfer}
\label{subsec:analysis_trnsf}

To study the transfer of energy from one band of modes to another, we consider shells of wavenumbers, $I_P$, defined by:
\begin{equation}
I_P = \{ (k_x, k_y, k_z ), ~ {\rm such~that} ~ 2^{P-1} k_f /\sqrt{2} \le |\mathbf{k}|
= \sqrt{k_x^2 + k_y^2 + k_z^2 } \le 2^{P-1} k_f \times \sqrt{2} \}
\label{eq:I_P}
\end{equation}
where $k_f$ is a fixed wavenumber, chosen here to be $k_f = 2.3$.
In elementary terms, the sets $I_P$ correspond to spherical shells in wavenumber space, around the wavenumber $K_P = 2^{P-1} k_f$, and of width $ \sim k_f$.
With the value of $k_f$ chosen here, the number of Fourier modes in the lowest shell, $I_1$, is $\sim 200$, which minimizes the effect of discretization.

Furthermore, we define the band-passed filter of the velocity field, $\mathbf{u}_P$, as:
\begin{equation}
\mathbf{u}_P ( \mathbf{x}, t) = \sum_{(k_x,k_y,k_z) \in I_P} \hat{\mathbf{u}}(k_x, k_y, k_z) \exp[  i ( k_x x + k_y y + k_z z) ]
\label{eq:def_u_p}
\end{equation}
In the same spirit, we define the filtered vorticity field $\mathbf{\omega}_P$ as the curl of $\mathbf{u}_P$, the filtered strain rate tensor $\mathbf{S}_P$ as the symmetric part of the derivative tensor of $\mathbf{u}_P$, the filtered kinetic energy as $ e_P(\mathbf{x}, t)  = \frac{1}{2}\textbf{u}_P (\mathbf{x}, t)^2$, the filtered the strain rate magnitude as $S_P = ||\textbf{S}_P||$, and the filtered dissipation rate, $D_P$, as $D_P=\frac{1}{2}\nu ||\textbf{S}_P(\textbf{x},t)||^2=\frac{1}{2}\nu S_P(\textbf{x},t)^2$.

To study the rate of energy transfer between scales,  we follow \cite{Alexakis:2005b,Mininni:2006}, and consider the rate of energy transfer $T_3(K, P, Q)$:
\begin{equation}
T_3(K, P, Q)= \int \mathcal{T}_3(K,P,Q, \textbf{x}) \, d \textbf{x} ~~~ {\rm with} ~~~ \mathcal{T}_3(K, P, Q, \textbf{x})= - \textbf{u}_K \cdot (\textbf{u}_P \cdot \nabla)\textbf{u}_Q \, .
   \label{eq:def_T3_int}
\end{equation}
This third-order correlator represents the energy transferred from shell $Q$ to shell $K$ due to the interaction with the velocity field in shell $P$. This operator can be derived from the Navier-Stokes equation, and we note that it does not give information on the amount of energy shells $Q$ and $K$ transfer to $P$, just on what $P$ transfers from $K$ to $Q$ \citep{Alexakis:2005b,Mininni:2006}.
Here, we do not restrict ourselves to the integral over the entire volume in Eq.~\ref{eq:def_T3_int}, but we also consider the integrand $\mathcal{T}_3(K,P,Q, \textbf{x})$.
In fact, the integral in Eq.~\ref{eq:def_T3_int} averages out potentially useful information contained in $\mathcal{T}_3(K,P,Q,\textbf{x})$. We also note that both $\mathcal{T}_3$ and $T_3$ generally depend on time. In the special case of statistically steady flows, upon which most earlier studies have focused, a time average is also performed in the definition of $T_3$. Here, we will examine both time-dependent and statistically stationary problems, in which case we will consider the temporally averaged value for $T_3$. 

Summing over the middle shells, $P$, one obtains the energy transfer from shell $Q$ to shell $K$ through the whole velocity field:

\begin{equation}
    \mathcal{T}_2(K,Q) = \sum_{P} \mathcal{T}_3(K,P,Q) = -\mathbf{u}_K \cdot( \mathbf{u} \cdot \nabla) \mathbf{u}_Q.
    \label{eq:def_T2_int}
\end{equation}

\noindent This operator can be integrated over space to obtain $T_2(K,Q)$, i.e. the energy transfer rate from shell $K$ to shell $Q$ in the entire volume. 

Finally, one can sum the operator $\mathcal{T}_{K,Q}$ or $T_{K,Q}$ over the shells $Q$ to obtain the rate of energy received by the shell $K$ from the whole velocity field, i.e.~all shells:

\begin{equation}
    \mathcal{T}_1(K) = \sum_Q \mathcal{T}_2(K,Q) = -\textbf{u}_K\cdot(\textbf{u}\cdot\nabla)\textbf{u} .
\end{equation}

\noindent Again, by integrating over the volume, one obtains $T_1(K)$. This parameter can then be used calculate the energy flux in the cascade $\Pi(k)$ as:

\begin{equation}
    \Pi(k) = -\sum_{K=0}^k T_1(K) 
    \label{eq:def_Pi}
\end{equation}

\noindent which expresses the energy flux through the shell $K$ \citep{Mininni:2006}.

In this manuscript, we will focus on the behaviour of $\mathcal{T}_2(K,Q)$, which we denote for simplicity as $\mathcal{T}_{K,Q}$. Similarly, we denote the averaged quantity, $T_2(K,Q)$ by $T_{K,Q}$. In an analogous manner, we express $\mathcal{T}_1(K)$ as $\mathcal{T}_K$, and so on. Furthermore, because in a turbulent flow, energy is ultimately transferred down to very large wavenumbers (very small scales) where it is dissipated by viscosity at a rate $\epsilon$, we systematically normalize the $\mathcal{T}$ and $T$ operators by dividing by the dissipation rate in the flow. 

In general, positive values of $\mathcal{T}_{K,Q}$ imply that energy is transferred from shell $Q$ to $K$, while negative values mean energy is transferred from $K$ to $Q$. All of the previously defined quantities have been determined numerically in our DNS and are presented in the analysis below.
We stress that one of the main novel aspects of our work consists of examining not only the integral
quantities $T_{K,Q}$, but also visualizing the field $\mathcal{T}_{K,Q} ( \mathbf{x}, t)$ in physical space. This quantity allows us to monitor where the energy transfer occurs in real space, even if the interpretation is not totally straightforward. Since the role of the mediating velocity component in the shell $P$ is averaged out, this could result in an oversimplification of the energy transfer process. Furthermore, the correlators $\mathcal{T}_{K,P,Q}$ and derived quantities are not Galillean invariant \citep{Aluie:2009}. However, in the examples considered here, the mean velocity is very small, or even exactly zero in the spectral simulations, which alleviates the issues mentioned in \cite{Aluie:2009}, and allows us to interpret these as good approximations for the physics of the energy transfer.

Finally, even if it is known that the $T_{K,P,Q}$ and $T_{K,Q}$ operators are antisymmetric, i.e.~$T_{K,P,Q}=-T_{Q,P,K}$ and $T_{K,Q}=-T_{Q,K}$, because energy is conserved by the triadic interactions, $\mathcal{T}_{K,Q}$ does not share this property as it is defined only up to a divergence. An elementary calculation leads us to the equality: $\mathcal{T}_{K,Q} + \mathcal{T}_{Q,K} = - \nabla \cdot [ \mathbf{u} (\mathbf{u}_K \cdot \mathbf{u}_Q) ]$. The divergence term integrates to $0$, which guarantees that the global balance $T_{K,Q} + T_{Q,K} = 0$, even if locally $\mathcal{T}_{K,Q} + \mathcal{T}_{Q,P}$ are not equal to zero.

\subsection{Details of the DNS}
\label{subsec:DNS}

The DNS for the two antiparallel vortex tubes were carried out using the energy conserving finite-difference code described in Appendix C of~\cite{McKeown:2018}. 
In essence, this is a centered, energy-conserving, second-order finite difference code which has a fractional time-stepping mechanism. 
The nonlinear terms are treated explicitly using a third-order Runge-Kutta scheme, while the viscous term is treated implicitly using a second-order Crank-Nicholson method. 
The simulations shown here were obtained in a cubic periodic domain of side length $\mathcal{L}$, with up to $540^3$ grid points for the circulation Reynolds number $Re_\Gamma=\Gamma/\nu=4500$, where $\Gamma$ is the circulation of the vortices and $\nu$ the kinematic viscosity of the fluid. 
The other parameter that defines the system is the ratio $a/b$, where $a$ is the radius of the vortex cores, and $b$ is the initial distance between the vortices. For this configuration, $a/b=0.4$, and $a/\mathcal{L}=0.06$.

Aside from the unforced anti-parallel tubes, which are left to freely evolve from initial conditions, we also simulate statistically stationary turbulence with two types of forcing: a random isotropic forcing \citep{Eswaran:1988}, with the forcing parameters chosen following \cite{Chouippe:2015}, and we also used a forcing inspired by \cite{Goto:2017} 
of the form:
\begin{equation}
 \mathbf{f} = \begin{bmatrix} 
- \sin(x) \, \cos(y) \\
 \cos(x) \, \sin(y) \\
 0
 \end{bmatrix} \, .
\label{eq:GotoF}
\end{equation}
This is a single-mode forcing which generates four counter-rotating vortex tubes, parallel to the $z-$axis and centered in the $(x,y)$ plane at $\pi/2 (2n+1, 2m + 1)$ with $n$, $m$ = $0$ or $1$. 

We also implement a spectral code to perform DNS of the Navier-Stokes equations, both to simulate homogeneous isotropic turbulence (HIT) and to investigate the configuration of two perpendicular vortex tubes~\citep{Ostilla:2021}. The code solves the Navier-Stokes equations in a triply periodic box $[- \pi , \pi]^3$, as described in \cite{Pumir:1994}. The DNS study of HIT was carried out using $512^3$ Fourier modes, at a Reynolds number $Re_\lambda = 210$. The quality of the spatial resolution can be judged by the product $k_{max} \eta $, where $k_{max}$ is the largest Fourier mode included in the simulation, and $\eta = (\nu^3/\epsilon)^{1/4}$, the Kolmogorov length. The value of $k_{max} \eta \approx 1.5$ shows that the small scales are satisfactorily resolved. We compare the results of the two codes to study HIT, and verified that the results presented in Section~\ref{sec:HIT} are robust. 

To simulate the interaction of two intially perpendicular vortex tubes, we insert two vortex tubes in the cubic domain, whose centerlines are given by $( x = y$, $z = d/2)$ and $(x = -y$, $z = -d/2)$, both with a circulation $\Gamma$, as detailed in~\cite{Ostilla:2021}.
The signs of the circulations are selected so that the two vortices close to $x = 0$ and $y = 0$ move in the direction  $x > 0$.
Numerically, we set $\Gamma = 1$ and $d = 0.9$. We do not force the flow and let the tubes freely evolve. We run the simulation at several resolutions, from $256^3$ up to $512^3$ grid points, to ensure spatial accuracy. The core size of the vortices was chosen to be $\sigma =0.4/\sqrt{2}$ slightly smaller than $d/2$. In this manuscript, we focus on the run at $Re_\Gamma=4000$.

For each flow configuration, the resulting flow parameters (i.e. the vorticity modulus, the strain modulus or energy transfer rate) are examined in the 3D visualization program Dragonfly (Object Research Systems).

\section{Characteristics of the energy transfer in homogeneous isotropic turbulence}
\label{sec:HIT}

In this section, we consider the energy transfer rate in the case of statistically stationary flows. We first re-analyze the transfer of energy for hydrodynamic turbulence at a moderate Taylor Reynolds number ($Re_\lambda = 210$), 
and show that it is consistent with turbulent flows at comparable $Re_\lambda$ in the literature~\citep{Mininni:2005,Domaradzki:2007, Eyink:2009}. 
This allows us to provide relevant points of benchmarking and comparison for the study of time-dependent flows investigated in the following sections. 

\begin{figure} 
 \centering
 \subfigure[]{
   \includegraphics[width=0.45\linewidth]{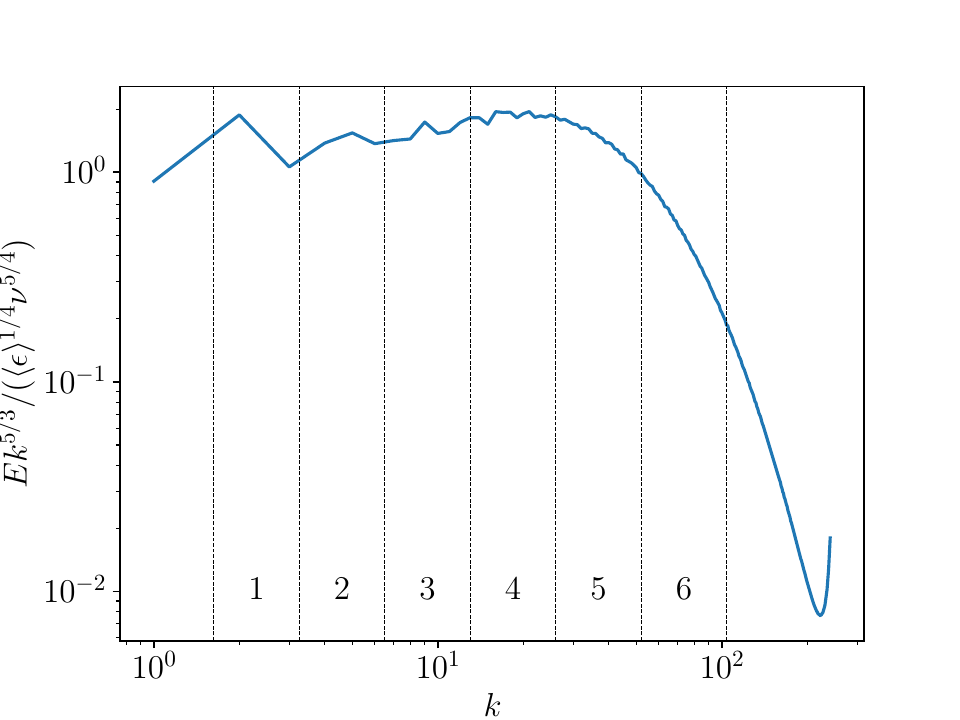}
}
\subfigure[]{
 \includegraphics[width=0.45\linewidth]{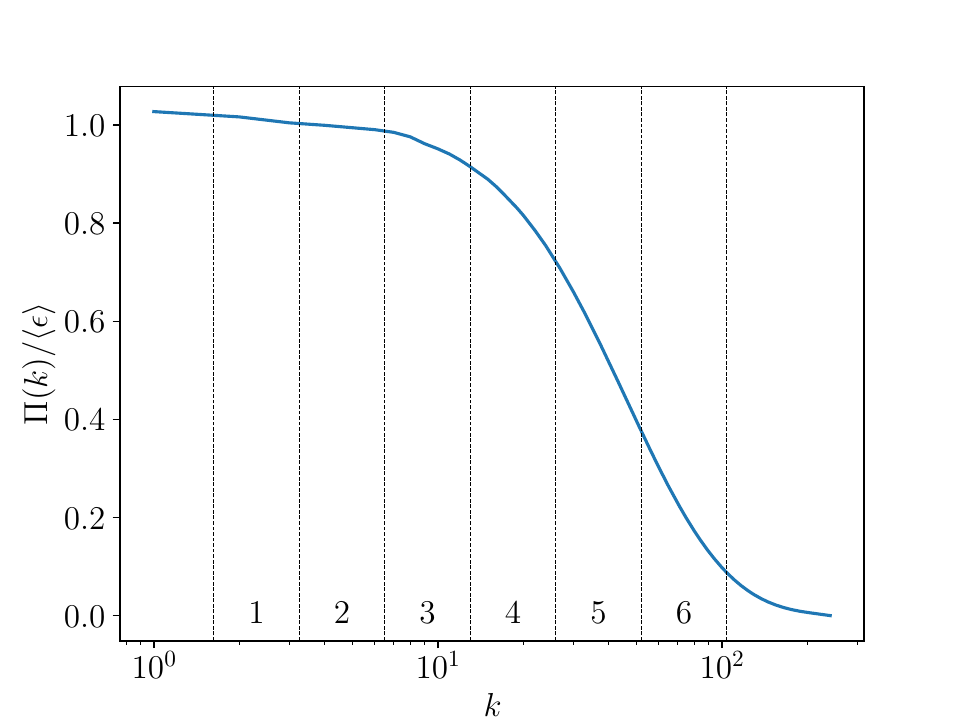}
 }
 \includegraphics[width=0.45\linewidth,trim={200 200 240 220},clip]{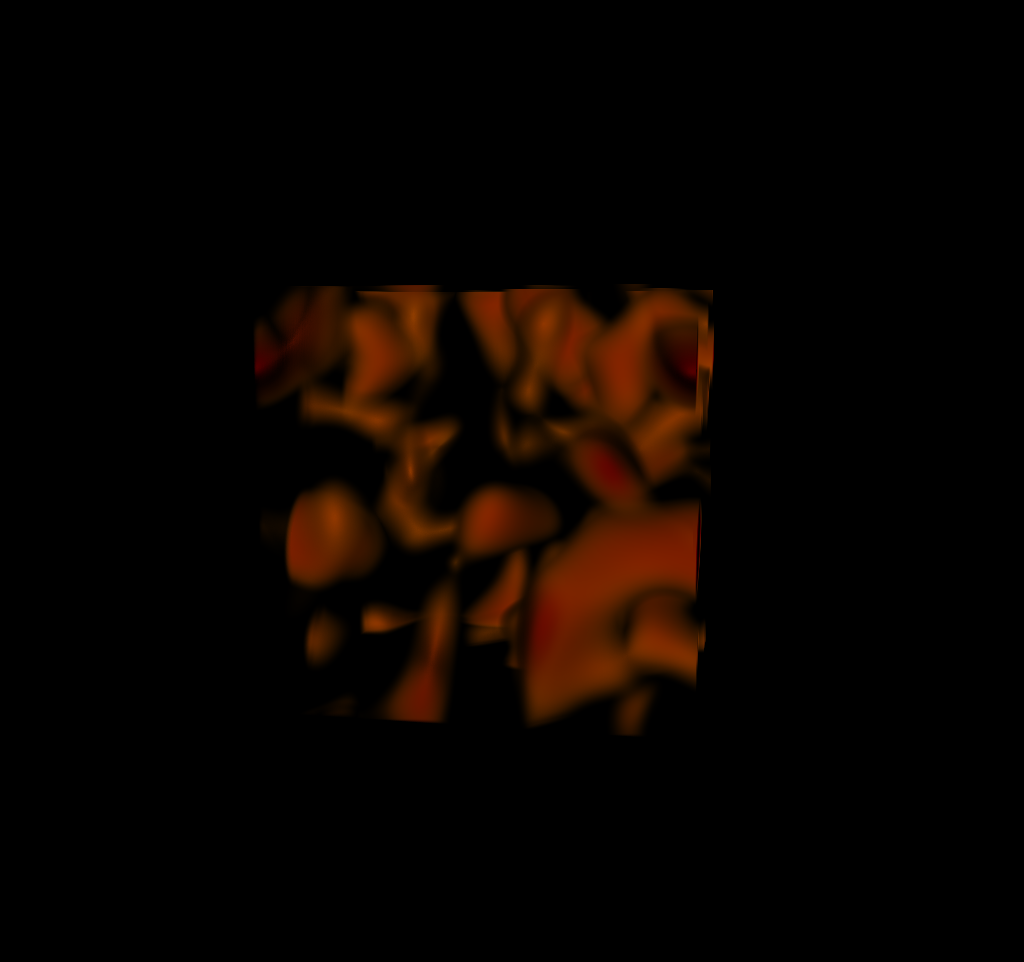}
 \includegraphics[width=0.45\linewidth,trim={200 200 240 220},clip]{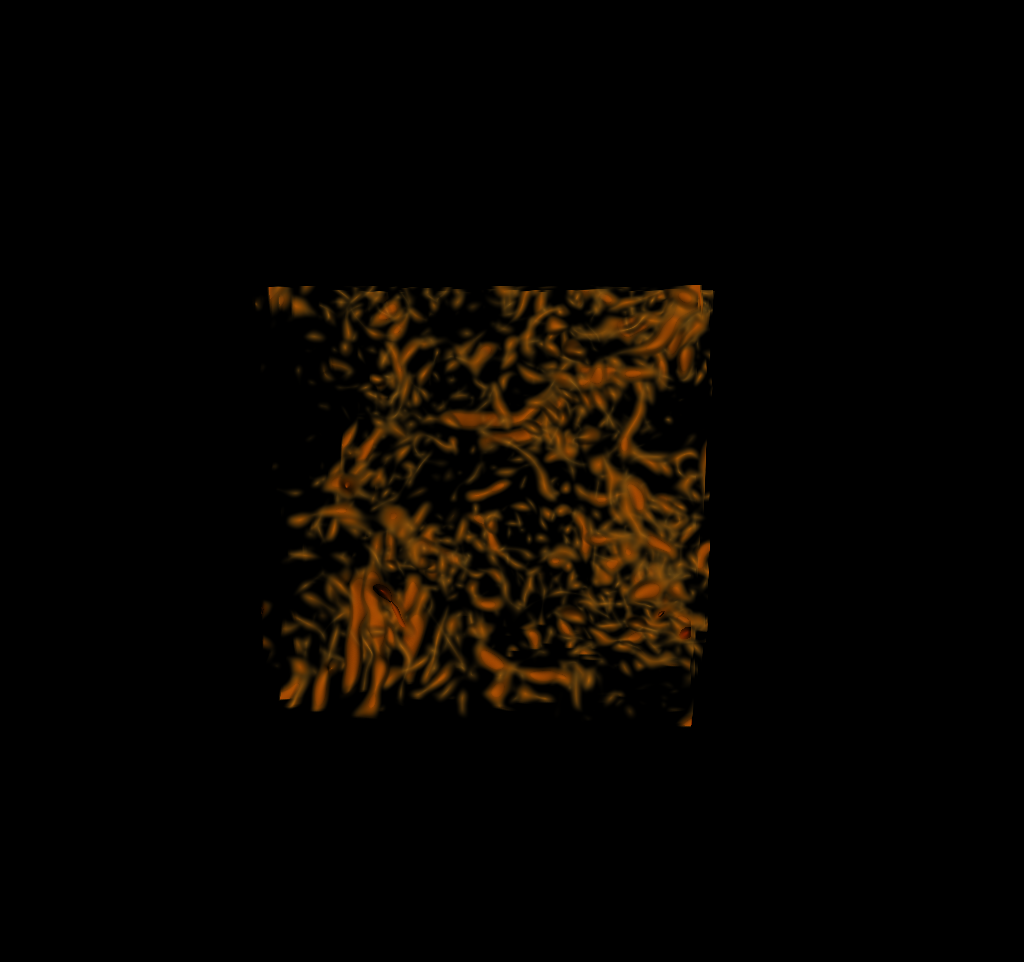}
 \includegraphics[width=0.45\linewidth,trim={200 200 240 220},clip]{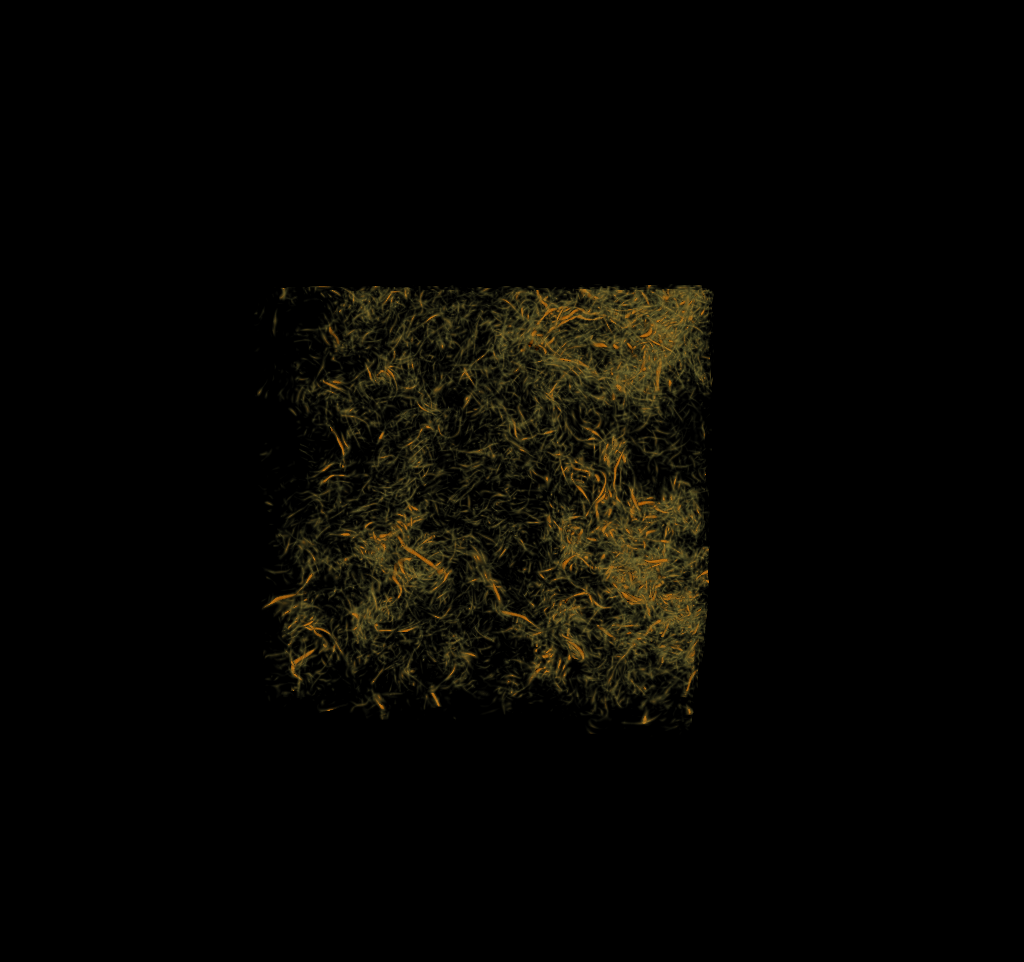}
 \includegraphics[width=0.45\linewidth,trim={200 200 240 220},clip]{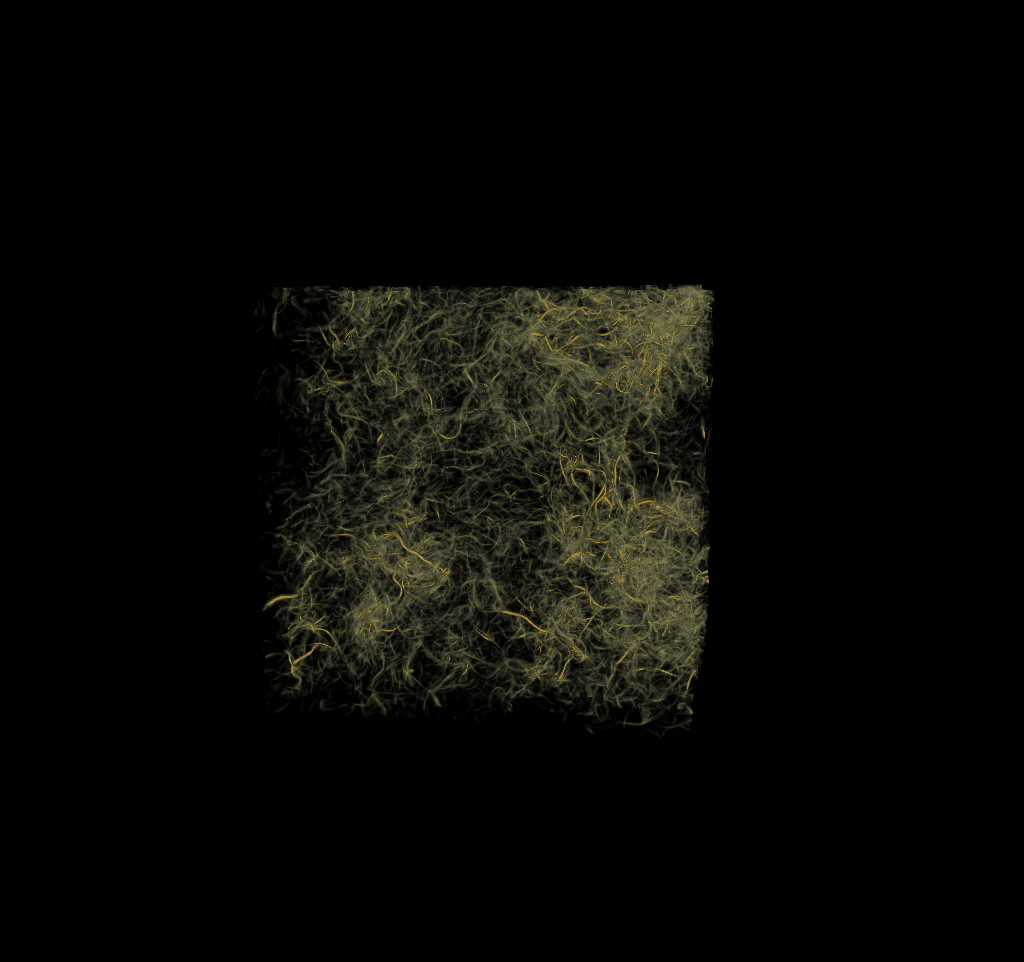}
  \caption{Statistics and visualization for homogeneous isotropic turbulence. (a) Shows the compensated energy spectra $Ek^{5/3}/(\langle\epsilon\rangle^{1/4}\nu^{5/4})$. (b) Shows the energy cascade rate $\Pi(k)/\langle \varepsilon \rangle$. (c-f) Show the vorticity filtered at scales $\omega_1$, $\omega_3$, $\omega_5$ and the unfiltered vorticity $\omega$, where regions of low vorticity have been made transparent.}
\label{fig:HIT_intro}
\end{figure}

To characterize the flow, we show the energy spectrum $E(k)$ and energy flux $\Pi(k)$ in Fig.~\ref{fig:HIT_intro}(a-b). We also indicate where the spectral shell bands start and end. With the value of $k_f = 2.3$ chosen to define the bands of wavenumbers for this flow, the range of inertial scales extends in wavenumber space from approximately the shell $I_2$, up to the shell $I_4$, which corresponds to the range $3.2 \lesssim k \lesssim 26.1$. The first shell is still largely affected by the forcing, and the two shells corresponding to the highest wavenumbers, i.e.~the shells $I_5$ and $I_6$, lose the $k^{-5/3}$ scaling and thereby correspond to the dissipative scales. To confirm this, we verified that the spatially-averaged dissipation rate $D_P$ is less than the transfer to small scales for $P \le 4$. This is reflected in $\Pi(k)$, normalized by $\langle \epsilon \rangle$, the time-averaged energy dissipation rate, being larger than 0.75 in the first four shells.

To further corroborate this, we show in Fig.~\ref{fig:HIT_intro}(c-f) the vorticity magnitude in the shells $I_1$, $I_3$ and $I_5$, as well as the unfiltered vorticity. We observe that despite the statistical homogeneity of the flow, at smaller scales, the vorticity loses homogeneity and becomes localized in any moment of time at preferential areas. In fact, the regions of low vorticity, correspond to regions of low kinetic energy in the fluid. Furthermore, in the small scales, the extremes of the distribution become more pronounced. This is reflected in a paling of the vorticity in the visualizations, from the brownish vortices of $\omega_1$ in Fig.\ref{fig:HIT_intro}(c) to the light vortices of $\omega_5$ in Fig.\ref{fig:HIT_intro}(f).

\begin{figure}
 \centering
\subfigure[]{ \includegraphics[width=0.45\linewidth]{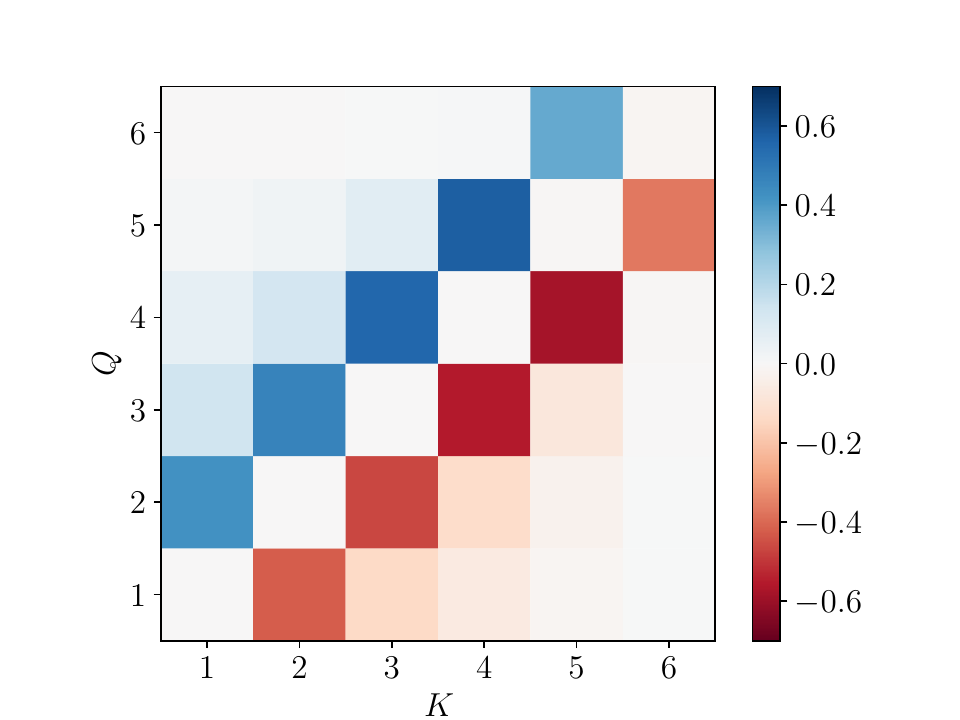}
}
\subfigure[]{
 \includegraphics[width=0.45\linewidth]{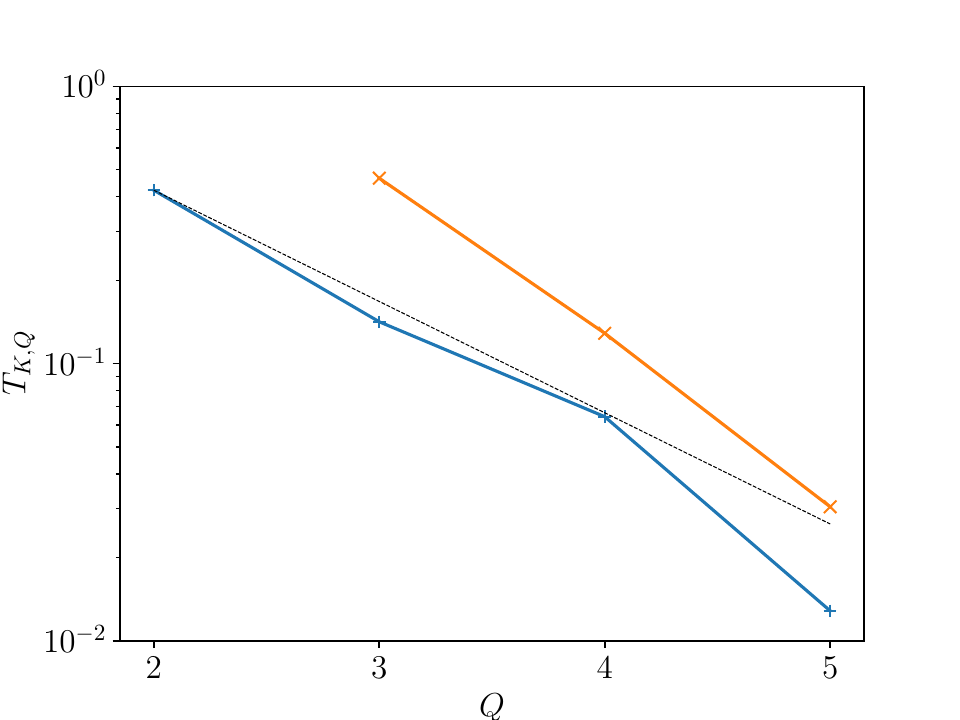} 
 }\\
 \subfigure[]{\includegraphics[width=0.45\linewidth]{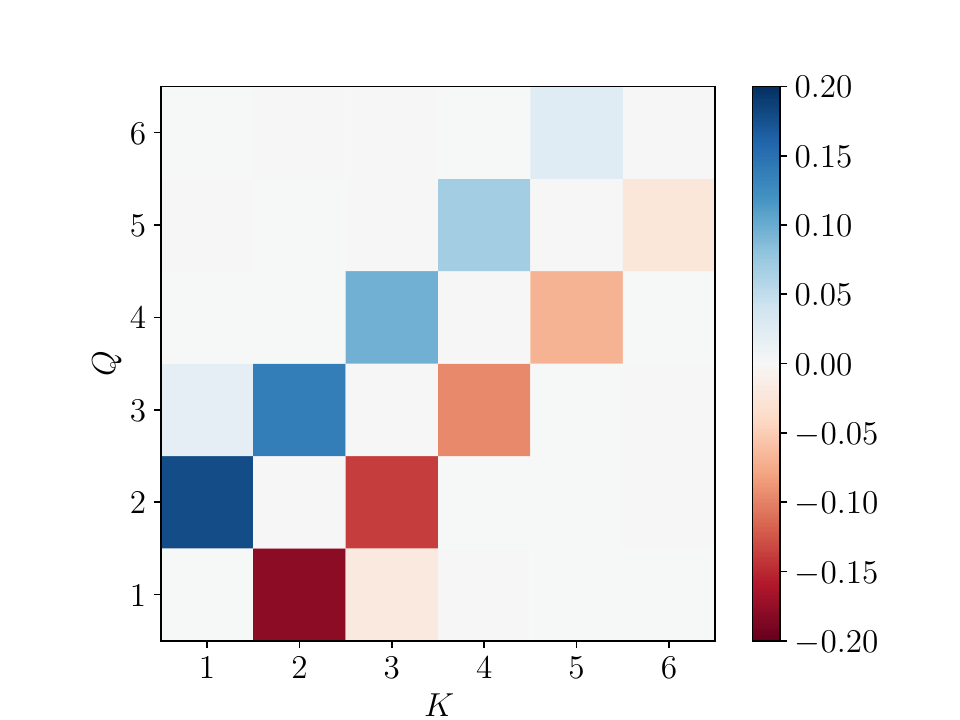}
 }
 \subfigure[]{\includegraphics[width=0.45\linewidth]{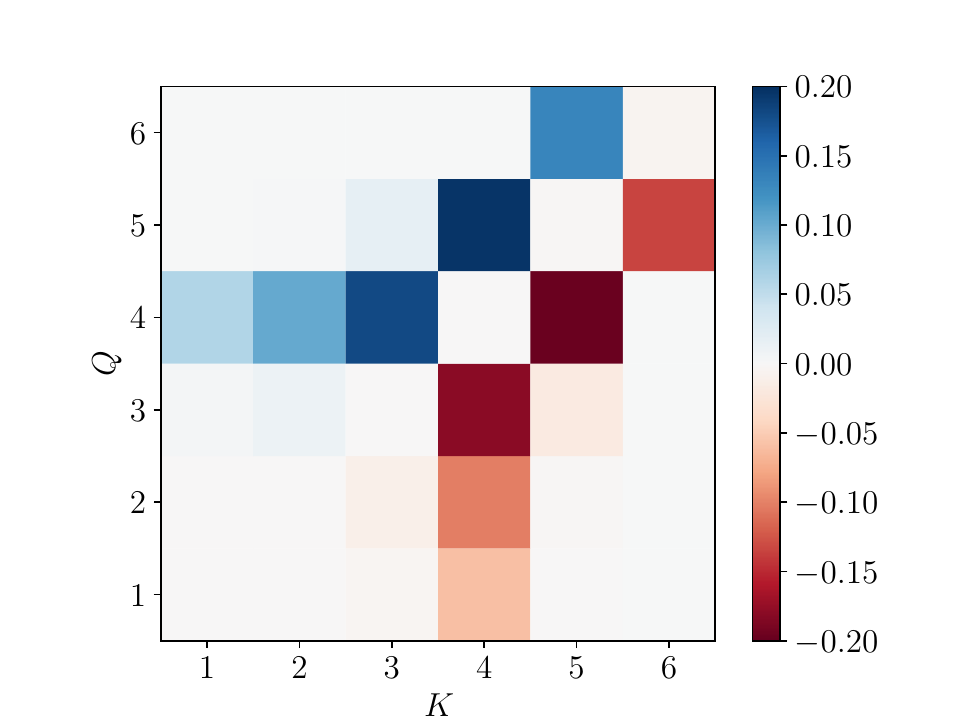}
 }
 \caption{Instantaneous integrated 2D energy transfer spectrum, $T_{K,Q}/\langle \epsilon \rangle$ in homogeneous isotropic turbulence (a)
at $Re_{\lambda} = 210$, where $k_f = 2.3$.
(b) shows the integrated transfer rate, $T_{K,Q}/\langle \epsilon \rangle$  from modes $K = 1$ (blue plus symbols) and from $K = 2$ (red cross symbols). The straight dashed line corresponds to an exponential decay: $\propto 2^{-4 Q/3}$.  The bottom panels show the triple correlator $T_{K,P,Q}/\langle\epsilon \rangle$ for (c) $P=2$ and (d) $P=4$. }
\label{fig:tkq_hit}
\end{figure}

Having characterized the base flow, Fig.~\ref{fig:tkq_hit}(a) shows $T_{K,Q}$. This quantity $T_{K,Q}$, defined by integrating Eq.~\eqref{eq:def_T2_int}, is calculated for the flow using $120$ different snapshots, uniformly distributed over a total duration of approximately one large eddy turnover time. We note that due to the statistically stationary nature of the flow and the integration of the energy transfer rate over all space, the values of $T_{K,Q}$ for each single snapshot do not vary considerably over this time from the averaged values over the $120$ snapshots, which we show here. We find that when $K < Q$, the rate of energy transfer is positive from bands of wavenumbers $I_K$ to bands of wavenumbers $I_Q$.
This corresponds to a transfer of energy toward small scales, a characteristic property of turbulent flows in three spatial dimensions. 
Interestingly, Fig.~\ref{fig:tkq_hit}(a) shows that the transfer of energy is mostly
concentrated around the diagonals, with a sharp decay away from the principal
diagonal.
The dominance of the $T_{K,Q}$ terms close to the diagonal, i.e.~from $I_K \rightarrow I_Q$ with $Q = K + 1$, is qualitatively consistent with the notion that energy transfer is mainly local in wavenumber space, as documented many times in turbulent flows with a comparable range of inertial scales
\citep{Domaradzki:1990,Zhou:1993,Alexakis:2005b,Domaradzki:2007,Mininni:2005,Verma:2018}. The transfer, however, also extends to larger modes: $I_K \rightarrow I_Q = I_{K+2}$ and $I_K \rightarrow I_Q = I_{K+3}$. The rate of energy transferred from a band of wavenumbers to a more remote band in Fourier space is a statistic that is affected by the precise real-space mechanism in which the energy is transported across scales \citep{Zhou:1993,Alexakis:2005b,Domaradzki:2009}. 
Several theoretical investigations of the decay of $T_{K,Q}$ as a function of $|K-Q|$ have led to the prediction of an exponential decay~\citep{Kraichnan:1966,Tennekes:1972,Aluie:2009}, corresponding to a power law decay in wavenumber space ($T_{K,Q}\sim k^{-4/3}$) since the center of the bands $I_Q$ varies exponentially, $\propto 2^Q$, see Eq.~\eqref{eq:I_P}.

We reproduce these scalings below to further showcase the locality of the energy transfer. Fig.~\ref{fig:tkq_hit}(b) shows the integrated energy transfer rate from $I_1$ (``+'' symbols), or $I_2$ (``$\times$'' symbols) to $I_Q = I_{K + n}$. Fig.~\ref{fig:tkq_hit}(b) is consistent with an exponential decay of $T_{K,Q}$ ($K=Q+n$) when $n$ increases. This relationship corresponds to a power law relating wavenumber and $T_{K,Q}$, since the center of the bands $I_Q$ also varies as $\propto 2^Q$. We also note that the decays seem to differ slightly between $K = 1$ and $K = 2$. We interpret this as an effect of the limited inertial range in the simulation. In Fig.~\ref{fig:tkq_hit}(b), we have also added straight line with a $2^{-4Q/3}$ scaling, which corresponds to a power law with an exponent $-4/3$ (again, since $k_Q$, the center of the band $I_Q$ $ \propto 2^{Q}$). This scaling agrees well with $T_{1,Q}$ and $T_{2,Q}$ when $Q\leq 4$. This power-law dependence, derived by \cite{Kraichnan:1966}, see also \cite{Aluie:2009}, has been found in other numerical simulations at similar or larger $Re_\lambda$ \citep{Domaradzki:2009,Eyink:2009}. The phenomenological approach by \cite{Tennekes:1972} reaches the same conclusion by analyzing the energy transfer rate between scales $I_K$ and $I_Q$ in terms of the deformation induced on a given vortex structure by the rate of strain at scale $I_Q$, and on the isotropy-restoring effect of strain at scale $I_K$. Such an approach would suggest a correlation between $\mathcal{T}_{K,Q}(\mathbf{x})$ and 
strain at scales $I_K$ and $I_Q$. 
Accordingly, we will use the exponential decay of $T_{K,Q}$ as a function of $|K-Q|$ to test for the adequacy of the time-dependent flows analyzed below in modeling the turbulent cascade.

A final point for comparison between HIT and the flows analyzed later is provided by the behaviour of the triple correlator $T_{K,P,Q}$, shown for $P=2$ and $P=4$ in Fig.~\ref{fig:tkq_hit}(c-d), respectively. For the $P=2$ mode, the triple correlator shows a similar behaviour to $T_{K,Q}$, predominantly concentrating transfer along the diagonal $I_K\to I_Q$ with $Q=K+1$. The rate of transfer to other modes, such as $Q=K+2$ and $Q=K+3$ is greatly reduced due to the character of the triadic interactions. We also observe a rapid decay in the influence of the $P=2$ mode for large values of $K$. For the $P=4$ triple correlator, shown in Fig.~\ref{fig:tkq_hit}(d), there is negligible transport of energy through the $I_4$ velocity shell if $K<4$, again due to the triadic nature of the interactions. Conversely, we observe an extended transport band for values of $K=4$ and $Q<K$ (and vice versa), indicating that this mode is responsible for the transfer of the off-diagonal energy modes. Once again, the dominant mode of transfer is $I_K\to I_Q=I_{K+1}$.

\begin{figure}
 \centering
 \includegraphics[width=0.43\linewidth]{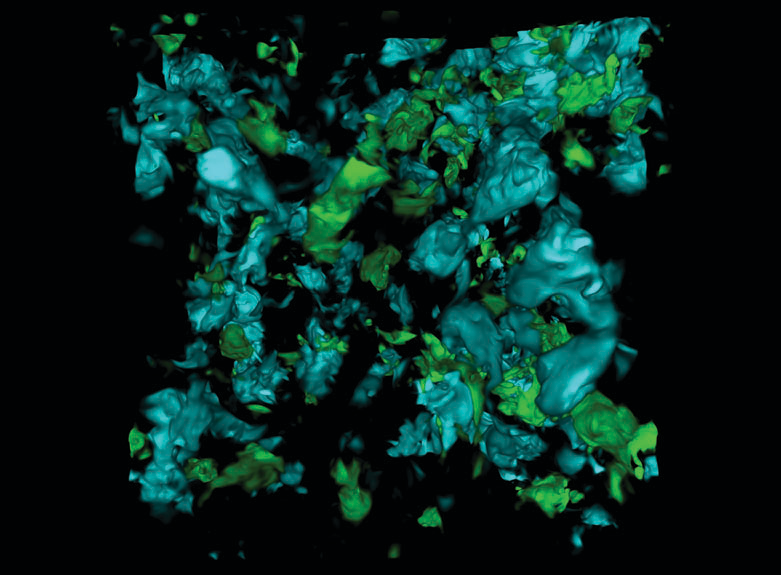}
 \includegraphics[width=0.43\linewidth]{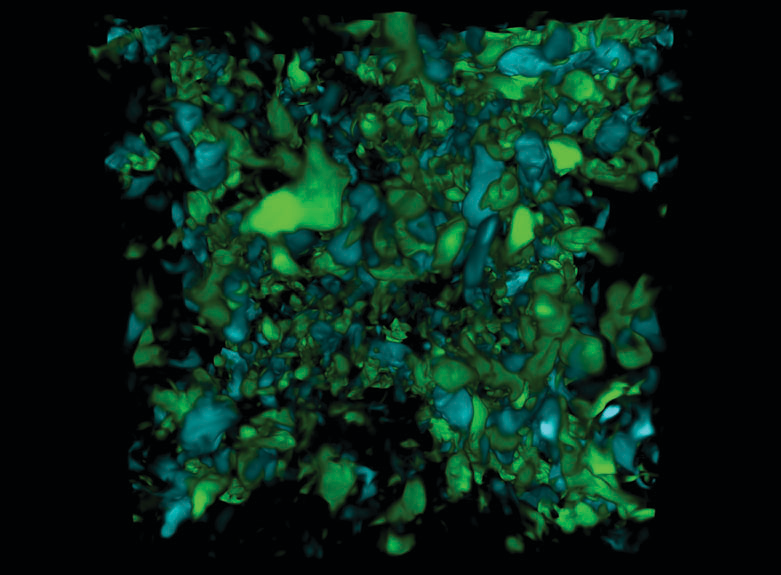}
 \includegraphics[width=0.43\linewidth]{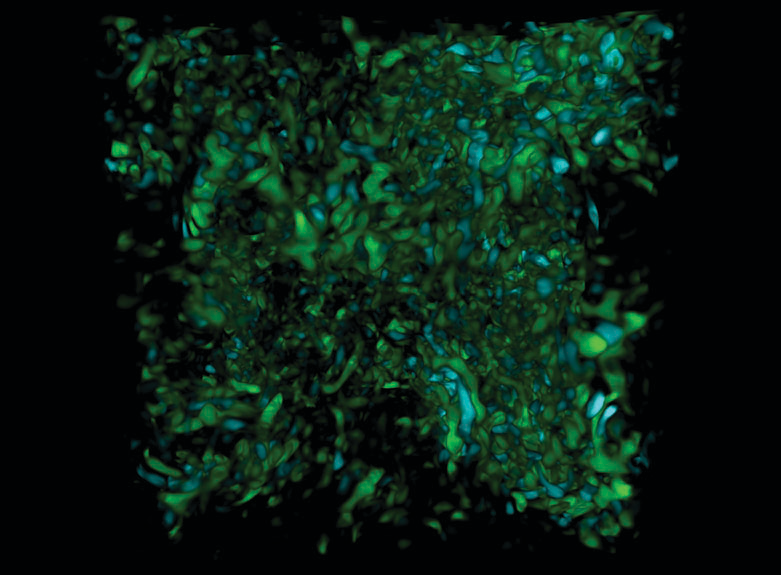}
 \includegraphics[width=0.43\linewidth]{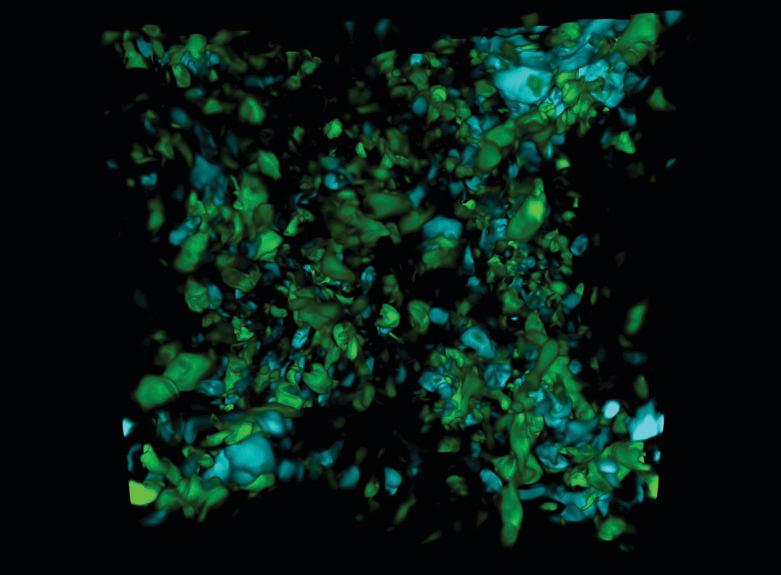}
 \includegraphics[width=0.43\linewidth]{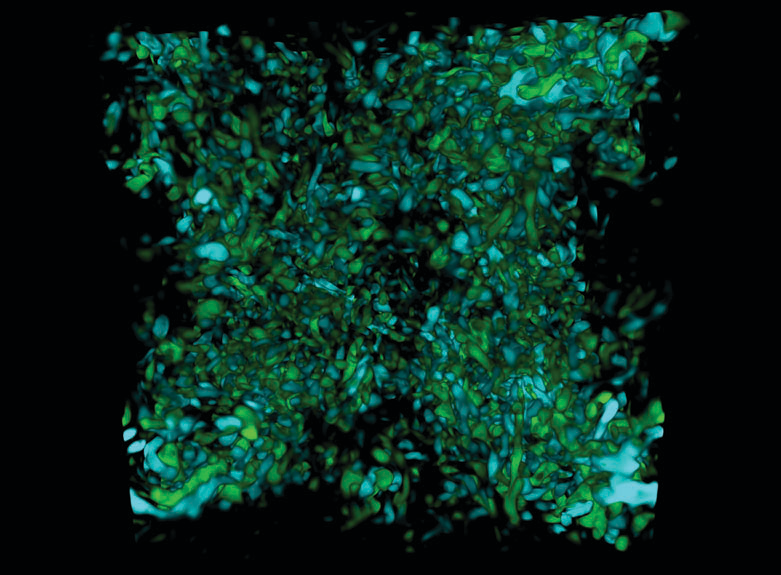}
 \includegraphics[width=0.43\linewidth]{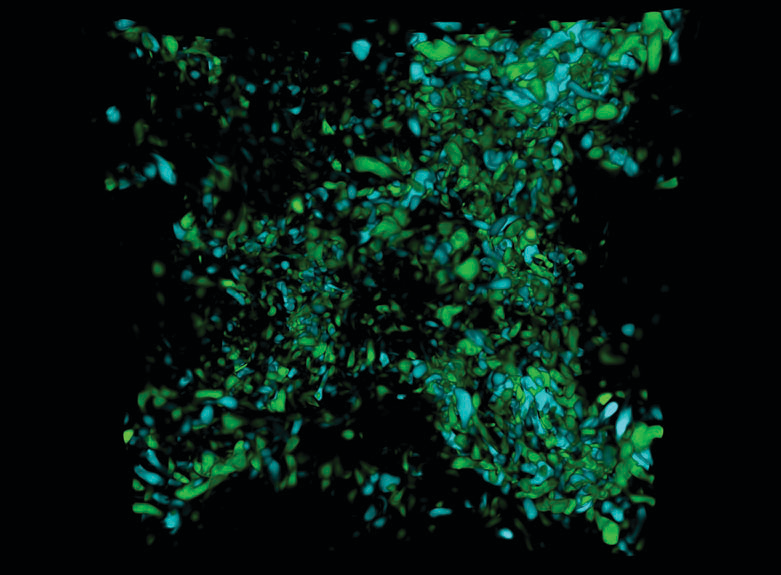}
  \caption{Instantaneous 2D energy transfer spectrum in homogeneous isotropic turbulence at $Re_{\lambda} = 210$, where $k_f = 2.3$. (a-f) Localized regions of intense energy transfer for (a) $\mathcal{T}_{1,2}$, (b) $\mathcal{T}_{1,3}$, (c) $\mathcal{T}_{1,4}$, (d) $\mathcal{T}_{2,3}$, (d) $\mathcal{T}_{2,4}$, (f) $\mathcal{T}_{3,4}$. Regions with a positive rate of energy transfer from larger to smaller scales are blue and regions with a negative rate of energy transfer from smaller to larger scales are green. }
\label{fig:tkq_vis_hit}
\end{figure}

As previously stated, an aim of this paper is to visualize the transfer of energy across scales directly by showing $\mathcal{T}_{K,Q}(\textbf{x})$ in real space. To this end, Fig.~\ref{fig:tkq_vis_hit} shows six snapshots of $\mathcal{T}_{K,Q}$ in HIT for different values of $K$ and $Q$, with the convention that blue regions correspond to a positive energy transfer towards small scales ($Q > K$), and green to a negative energy transfer. These are localized in the same regions of intense vortical activity visualized in Fig.~\ref{fig:HIT_intro}, especially as the scales become smaller. However, the overall picture of the cascade is hard to observe from these snapshots, especially for finer scales.  By calculating the Pearson correlation coefficients, we can make some more quantitative observations. The Pearson correlation coefficient between $\mathcal{T}_{K,Q}$ and $\omega_Q$, $\omega_K$, $S_Q$ and $S_K$ leads us to values which are consistently negative, but with a magnitude of $\sim 10^{-2}$, i.e.~very close to zero. If the correlations are made with respect to the absolute value of $|\mathcal{T}_{K,Q}|$ instead, the correlation coefficients increase to the range of $0.2-0.3$, though slightly higher for strain than for vorticity. This indicates that energy transfers occurs predominantly in regions of high vorticity, although the transfer can go in both directions. This is manifested by the absence of
correlation with between the signed quantity $\mathcal{T}_{K,Q}$ and vorticity. 
Additionally, visualizing the regions of intense vorticity and strain, filtered around a shell $I_K$, does not reveal any qualitatively strong correlation between the two fields. This impression is also confirmed by the weak values of the Pearson correlation coefficients we found, at most of order $0.3$.

These statistics and visualizations confirm what is already known, i.e.~that the turbulent cascade is a process involving the interaction of multiple scales. 
In particular, the lack of correlation between $\mathcal{T}_{K,Q}(\mathbf{x})$ and $S_{K} (\mathbf{x})$, $S_Q (\mathbf{x})$, or the vorticity components $\omega_K(\mathbf{x})$ and $\omega_Q(\mathbf{x})$, indicates that the enticing energy cascade proposed by~\cite{Tennekes:1972}, resting on a description of the action of strain and vorticity at neighbouring scales, is oversimplified; it does, however, provide the correct prediction for the qualitative scaling of the integrated rate of energy transfer, $T_{K,Q}$. In the following sections, we consider simpler flows, in order to obtain a clearer visualization of the energy transport mechanisms in real space. The analysis presented above will be used to set comparison benchmarks for the statistical metrics of these simpler flows, which will indicate how representative these flows are of the general turbulent cascade observed in HIT. 

\section{Time-dependent flows: energy transfer in interacting vortex tubes. }
\label{sec:t-dependent-Fourier}

\begin{figure}
 \centering
 \subfigure[]{\includegraphics[width=0.48\linewidth]{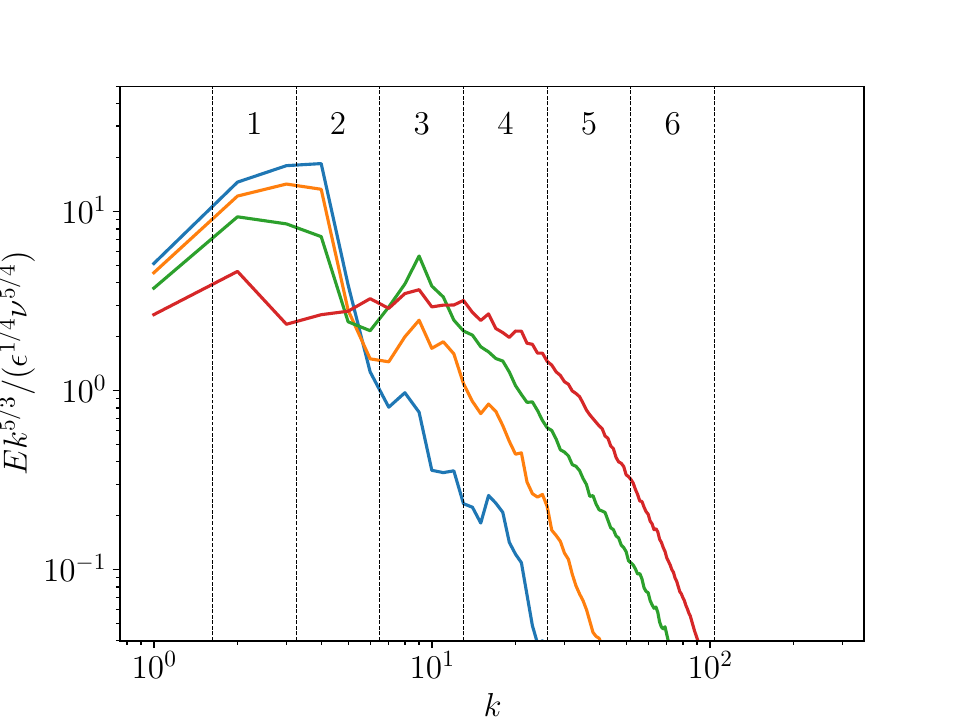}
 }
 \subfigure[]{\includegraphics[width=0.48\linewidth]{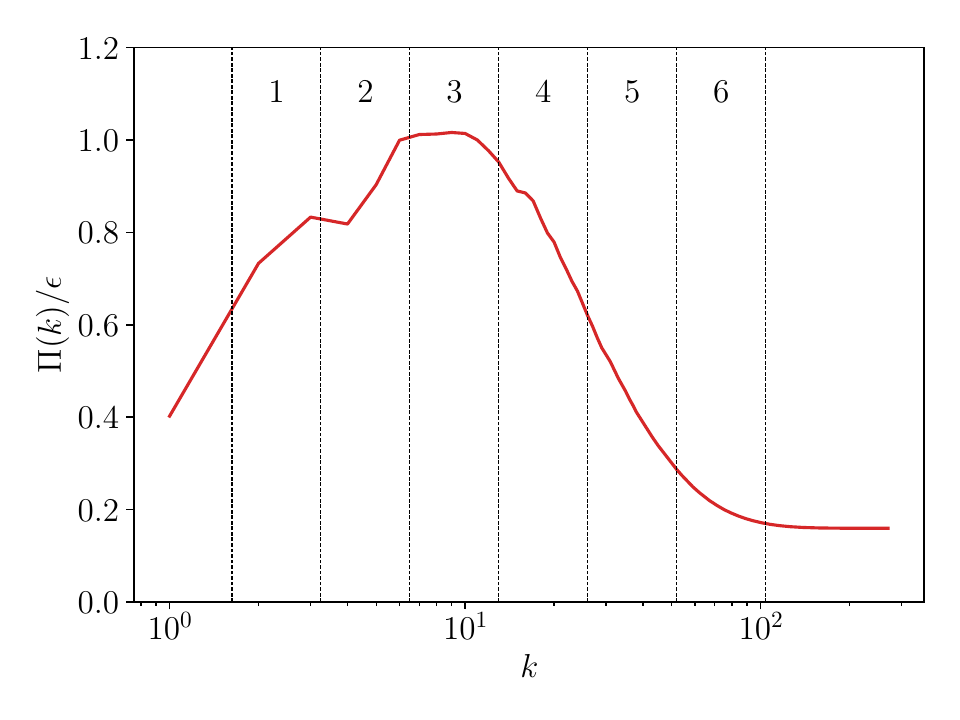}
 }
 \caption{(a) Instantaneous compensated energy spectra for anti-parallel tubes at times $t=50.0$ (blue), $t=60.3$ (orange), $t=68.3$ (green) and $t=82.3$ (red), which are analyzed below. The $k^{-5/3}$ spectra can be seen to emerge. (b) Energy flux spectrum, $\Pi(k)$, at $t=82.3$. The circulation Reynolds number is $Re_\Gamma=4500$.  }
\label{fig:tubes_spectra}
\end{figure}

In our previous work, we demonstrated that the interaction between antiparallel vortex tubes leads to an energy cascade composed of discrete and iterative events which were  isolated and visualized \citep{McKeown:2020}. These dynamics result in the formation of a transient $E(k)\sim k^{-5/3}$ energy spectrum, as shown in Fig.~\ref{fig:tubes_spectra}(a). We also show the instantaneous energy flux $\Pi(k)$ in Fig.~\ref{fig:tubes_spectra}(b) during the turbulent breakdown, divided by the dissipation rate at the corresponding time, $\epsilon$. We now visualize the transfer of energy during the flow evolution, from smooth antiparallel tubes to a turbulent cloud and  compare with the results obtained for HIT to assess how well this breakdown reproduces the canonical turbulent cascade.

The Reynolds number $Re_\Gamma$ is defined here as the dimensionless ratio between the circulation of the tubes at $ t= 0$, $\Gamma$, and the kinematic viscosity: $Re_\Gamma \equiv \Gamma/\nu$.
As shown by \cite{McKeown:2020}, at high Reynolds numbers, $Re_\Gamma \gtrsim 4000$, initially anti-parallel vortex tubes are subjected to an elliptical instability, leading to the formation of anti-parallel tubes, perpendicular to the original vortices. The pattern reproduces itself in a cascade-like manner, eventually giving rise to turbulence. In the following, we distinguish the two phases of the evolution and characterize the corresponding transfers of energy between bands of wavenumbers. We analyze the run at $Re_\Gamma = 4500$ from~\cite{McKeown:2020}. The time, $t$, is non-dimensionalized using $b^2/\Gamma$, where $b$ is the initial tube separation, as done in~\cite{McKeown:2020}.

\subsection{Early stage: development of the instability}
\label{subsec:par_early}

\begin{figure}
\centering
  \includegraphics[width=0.3\linewidth]{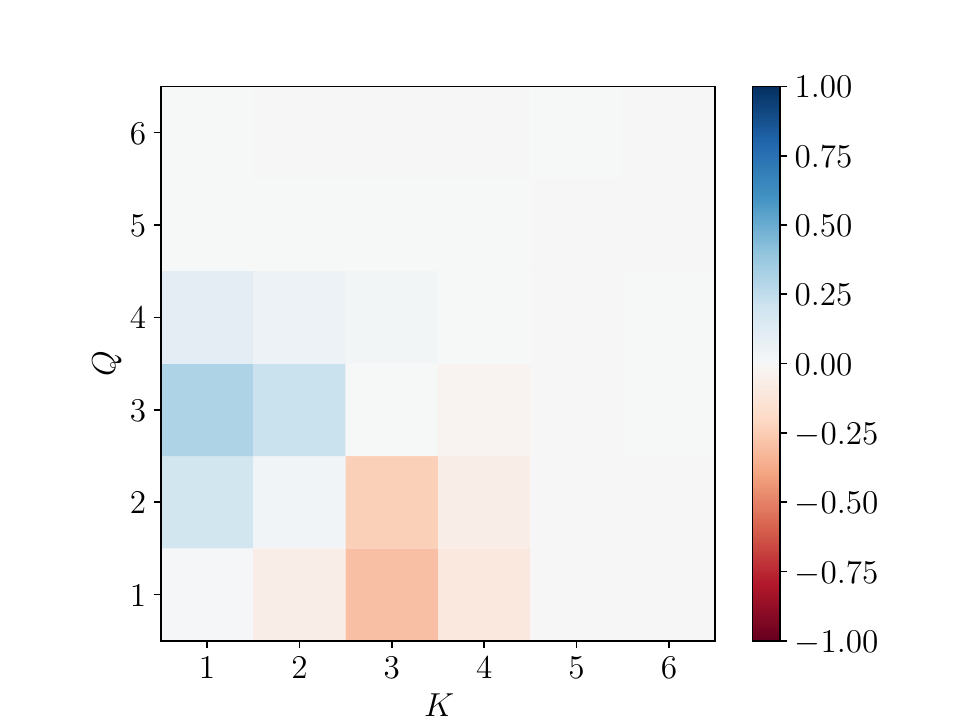}
    \includegraphics[width=0.3\linewidth]{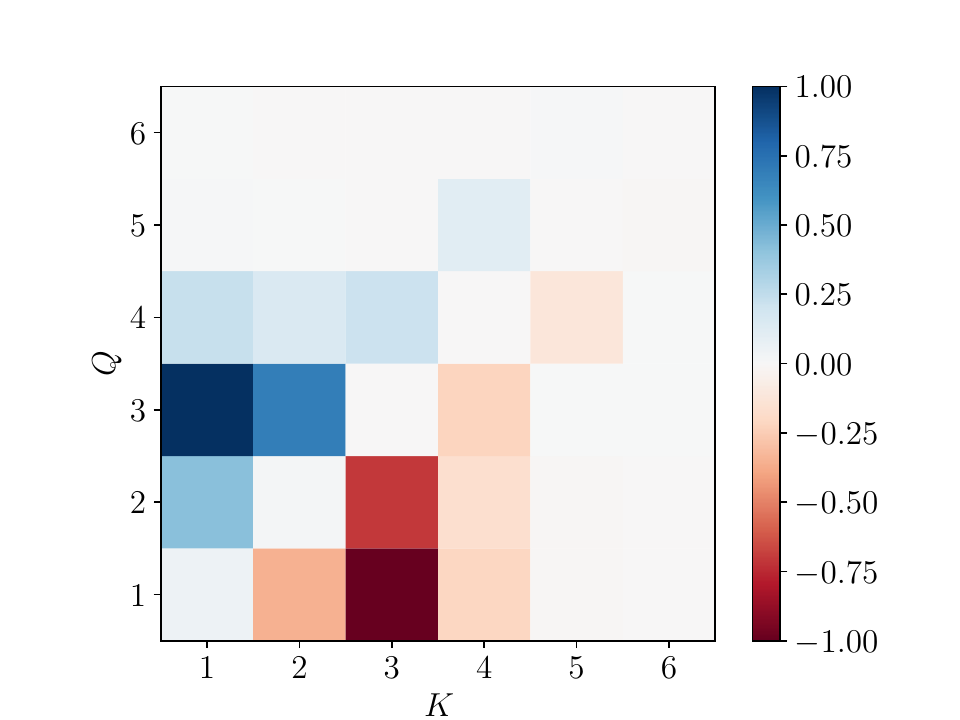}
      \includegraphics[width=0.3\linewidth]{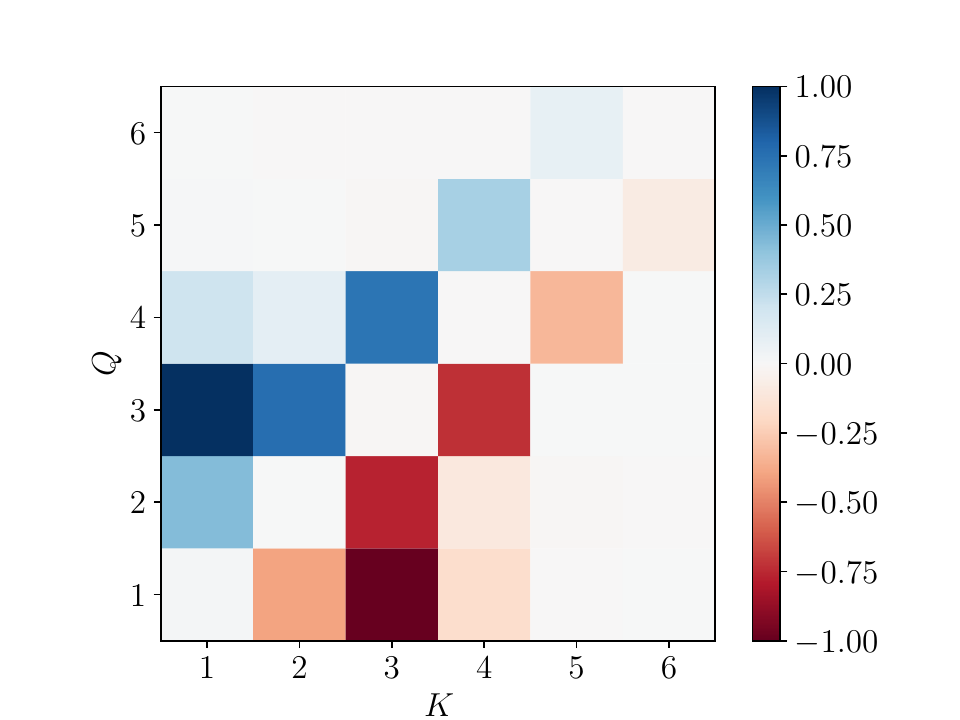}\\
  \includegraphics[width=0.9\linewidth, trim={0 0 0 4cm}, clip]{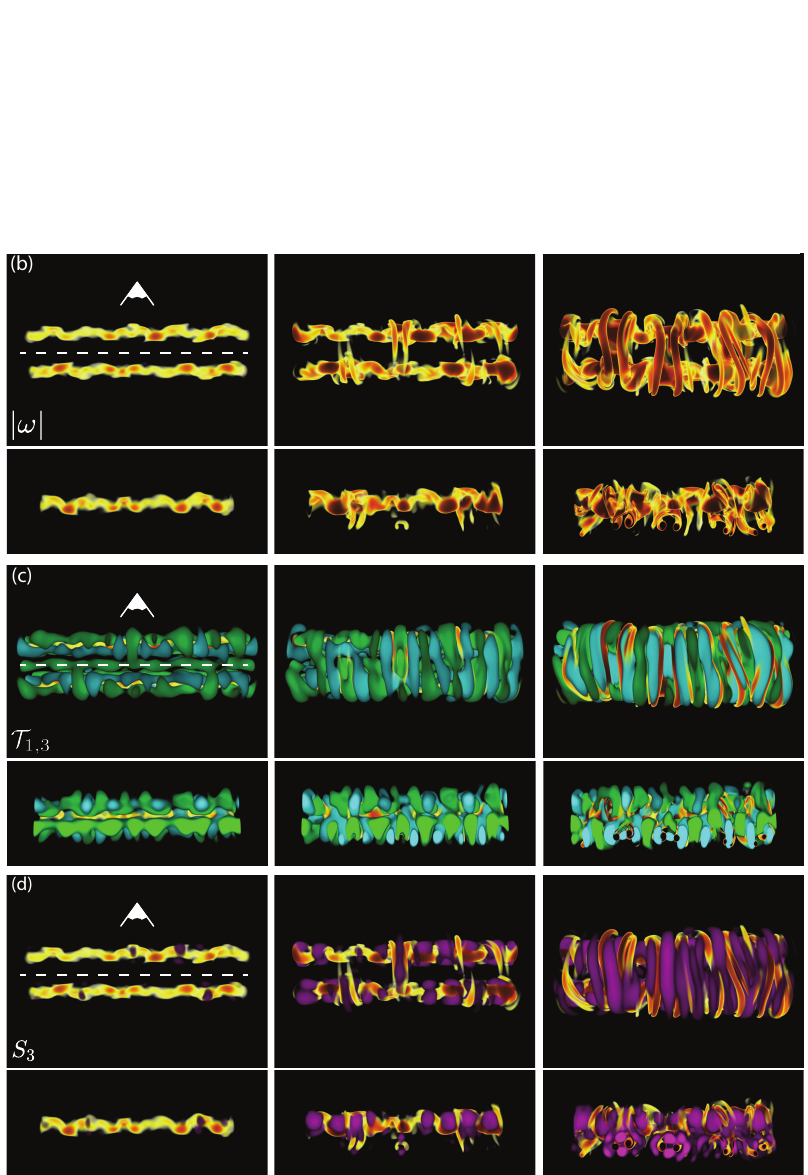}
  \caption{Energy transfer for two vortex tubes originally 
antiparallel, at times  
$t = 50.0$ (left column), when the first instability begins to develop;
$t = 60.3$ (central column), as the transverse filaments grow, 
and $t = 68.3$ (right column), as the transverse filaments begin to interact with each other.
(a) Integrated 2D energy transfer spectrum, normalized by the instantaneous dissipation rate, $T_{K,Q}/\epsilon$, where $\epsilon$ is the dissipation rate at the corresponding time.
(b) Front view (top) and cross-sectional top view (bottom) through the dashed white line showing the evolution of the vorticity modulus, as the elliptical instability develops.
(c) Regions of intense positive energy transfer to smaller scales,
from $I_1$ to $I_3$ (blue), and of intense negative energy transfer (to 
larger scales) from $I_3$ to $I_1$ (green). The presence of a strong
energy transfer correlates with the formation of perpendicular secondary vortex tubes.
(d) Filtered rate of strain magnitude at the $I_3$ band (magenta). Regions of intense strain correlate  strongly between interacting secondary filaments.
}
\label{fig:3}
\end{figure}

Fig.~\ref{fig:3} illustrates the evolution of the two tubes during the phase where the dissipation rate strongly increases, before a turbulent, $k^{-5/3}$ energy spectrum develops. Fig.~\ref{fig:3}(a) shows the integrated energy transfer rate, $T_{K,Q}$ in Fourier space,  whereas Fig.~\ref{fig:3}(b) shows the development of the instabilities through volumetric visualizations of the vorticity modulus, $\omega(\mathbf{x})$. 
At the earliest time shown ($t = 50.0$), the instability begins to develop, as indicated by the perturbations in the vortex cores, 
and little energy transfer occurs across scales. At the intermediate time, 
$t = 60.3$, the most pronounced transfer is between the band $I_1$ and the band $I_3$. This corresponds to a wavelength equal to roughly $1/8$ the size of the computational box, which corresponds to the perturbation wavelength from the instability. At a later time $t = 68.3$, the energy transfer also encompasses other pairs of modes at higher wavenumbers, which all correspond to energy transfer toward smaller scales. These transpire in adjacent shells, i.e. $I_K\to I_Q$ with $Q =K+1$, similar to what was observed in the previous section for the case of homogeneous isotropic turbulence.

As the most intense energy transfer over the time interval illustrated in Fig.~\ref{fig:3}(a) occurs between $I_1$ and $I_3$, panel (c) shows a visualization of $\mathcal{T}_{1,3}$, with regions of intense positive (negative) energy transfer from $I_1$ to $I_3$ in blue (green). As the flow develops, an alternating pattern of regions of positive and negative energy transfer emerges perpendicular to the interacting tubes. The transfer of energy is predominantly negative (green) between pairs of interacting vortex tubes. Since the quantities $T_{K,Q}$ in Fig.~\ref{fig:3}(a) are scaled by the instantaneous energy dissipation, which remains relatively weak up to $t = 60.3$, this implies that the exchange of energy is not particularly large at this instant. Physically, the growth of energy in the band of wavenumbers $I_3$ mostly arises from this initial instability. At later stages ($t = 68.3$), however, a positive energy transfer between the tubes (blue regions in Fig.~\ref{fig:3}(c) overwhelms the negative energy transfer (green). This observation indicates a qualitative change in the interaction between the two tubes, from a regime governed by the growth of an instability, to a regime with strong nonlinear interactions and energy transfer.

It is interesting to note that the rate of strain filtered in the $I_3$ band, and visualized in magenta in Fig.~\ref{fig:3}(d), closely follows the local amplification of positive and negative energy transfer seen in Fig.~\ref{fig:3}(c). In fact, at the early times shown ($t = 50.0$ and $t = 60.3$), the development of intense strain regions precedes the formation of intense tubes, clearly visible at $t = 68.3$. 

The qualitatively strong correlations between the structures of intense regions of $S_Q$ and $\omega_Q$, clearly visible in Fig.~\ref{fig:3}, are confirmed by evaluating the Pearson correlation coefficients between $\omega_Q$ and $S_Q$. We find a value of the correlation coefficient of order 0.7-0.8, much stronger than what was found for HIT ($\approx 0.3$). The regions of intense transfer, $|\mathcal{T}_{K,Q}|$ also exhibit a stronger degree of correlation with strain or vorticity, filtered in the wavenumber band $I_Q$ or $I_K$, with a Pearson coefficient of $\approx 0.5$. Furthermore, at $t=60.3$ and $t=68.3$, we find a correlation coefficient of $\approx -0.15$ between $\mathcal{T}_{1,3}$ and $\omega_1$, as well as between $\mathcal{T}_{1,3}$ and the variables $\omega_3$, $S_1$ and $S_3$, unlike the practically zero correlation coefficients found in HIT. {This indicates that transfer across this pair of shells is comparatively more associated with a forward transfer of energy which happens in regions of high vorticity/strain than what is present in a HIT flow. A consequence of this observation is that energy transfer can be examined locally, at these times, through the interactions of discrete flow structures.
However, the flow is not yet comparable to homogeneous isotropic turbulence, as the energy spectrum differs from the classical $k^{-5/3}$-regime, and the structure of the $T_{K,Q}$ spectrum also substantially differs from what is observed in HIT flows, as examined in Sec.~\ref{sec:HIT}.

\subsection{Later stage: the turbulent regime}
\label{subsec:par_late}

\begin{figure}
 \centering
 \subfigure[]{
 \includegraphics[width=0.45\linewidth]{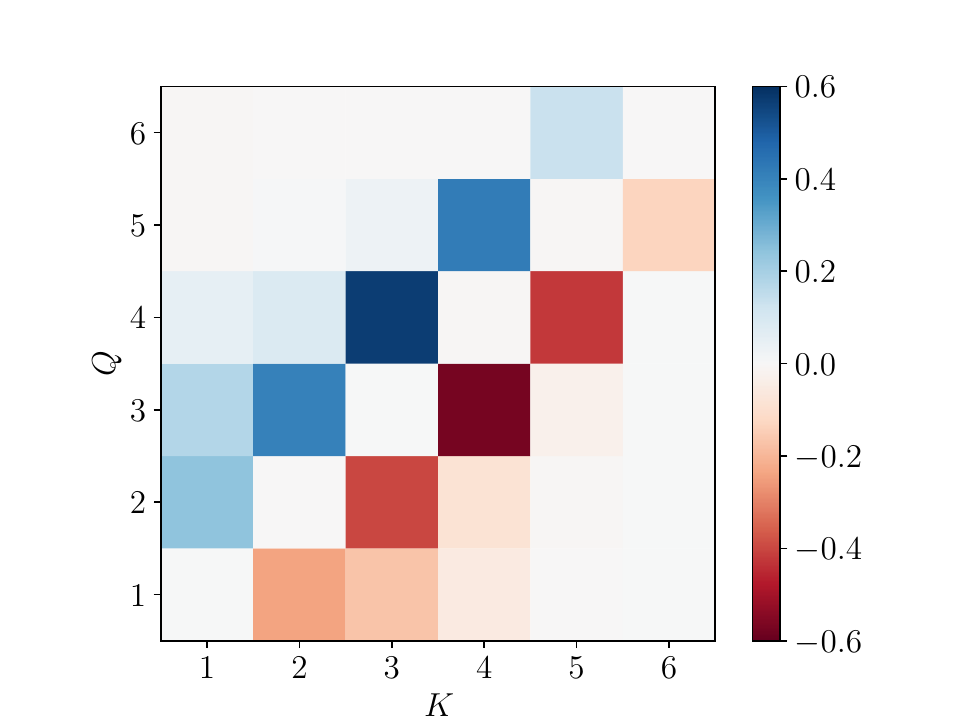}
 }
 \subfigure[]{\includegraphics[width=0.45\linewidth]{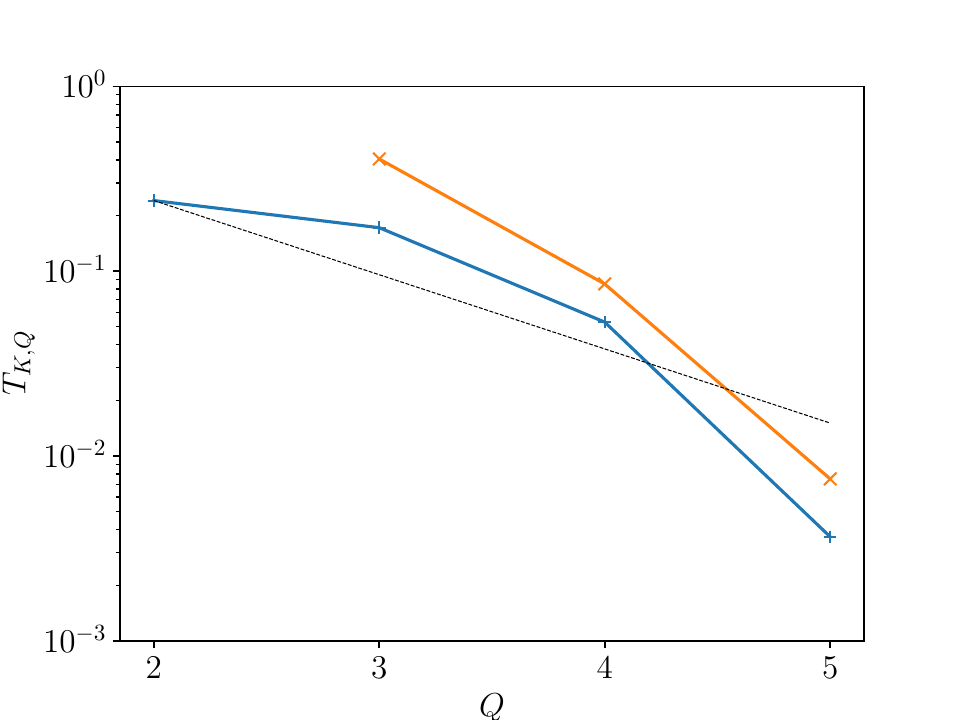}
 }\\
 \subfigure[]{\includegraphics[width=0.45\linewidth]{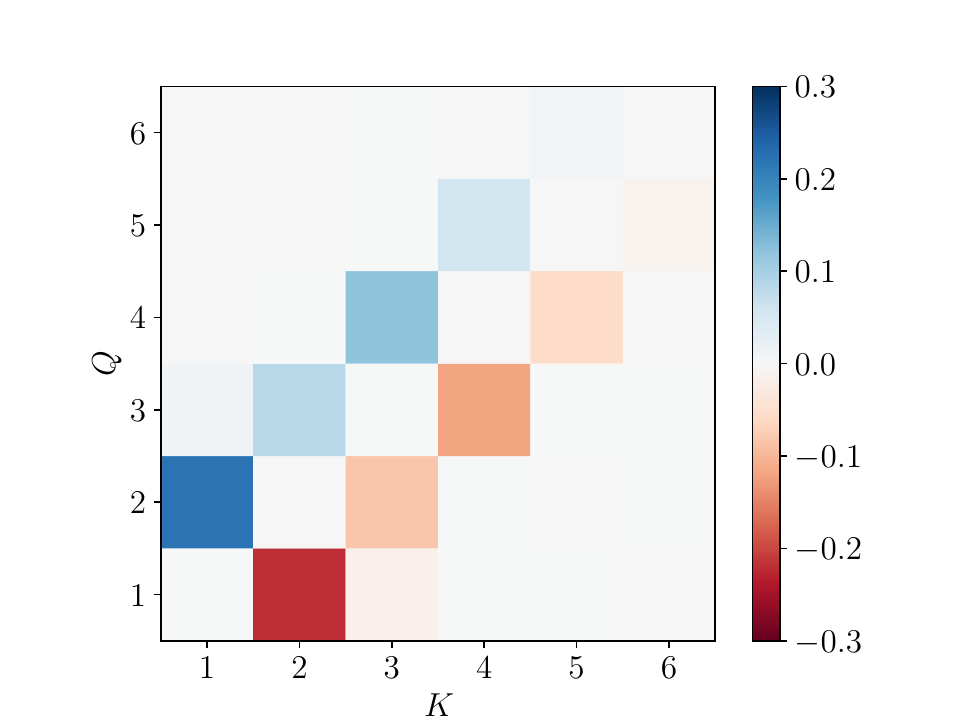}
 }
 \subfigure[]{\includegraphics[width=0.45\linewidth]{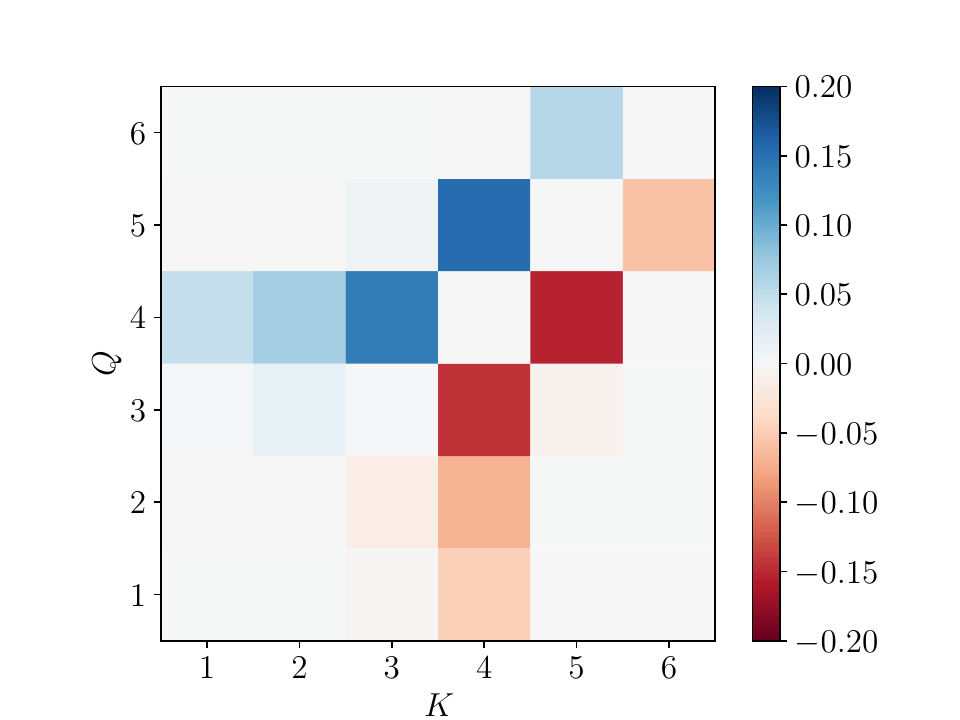}
 }
 \caption{(a) Instantaneous integrated energy transfer spectrum, $T_{K,Q}/\epsilon$, for the two interacting anti-parallel tubes at the time, $t=82.3$, where the dissipation rate, $\epsilon$, reaches its maximum vlaue. (b) shows the integrated transfer from modes $K = 1$ (blue plus symbols) and from $K = 2$ (orange cross symbols). The straight dashed line corresponds to an exponential decay: $\propto 2^{-4 Q/3}$. The bottom panels show the triple correlator $T_{K,P,Q}/\epsilon$ for (c) $P=2$ and (d) $P=4$. }
\label{fig:tkq_tubes_peak_dissipation}
\end{figure}

\begin{figure}
\centering
  \includegraphics[width=0.4\linewidth]{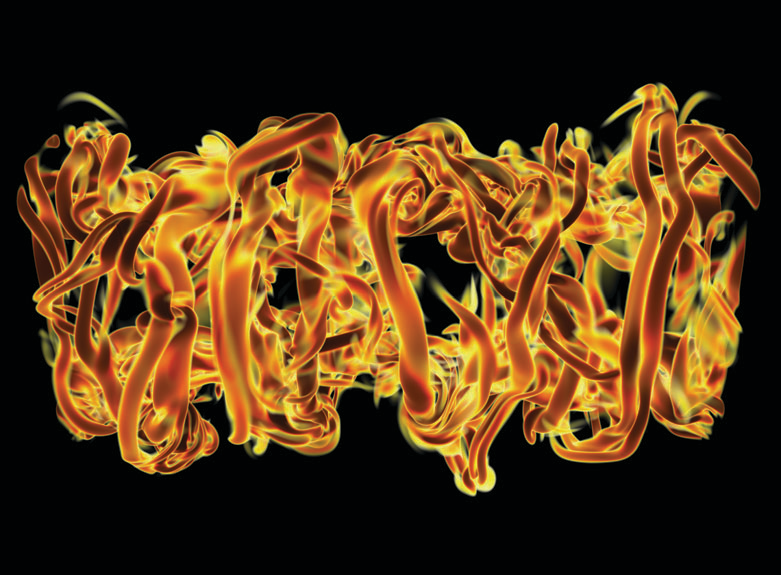}\\
  \includegraphics[width=0.4\linewidth]{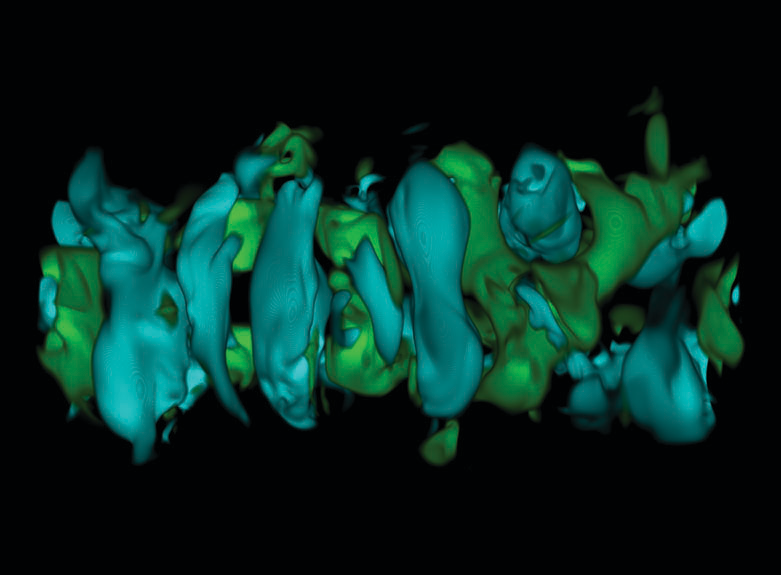}
  \includegraphics[width=0.4\linewidth]{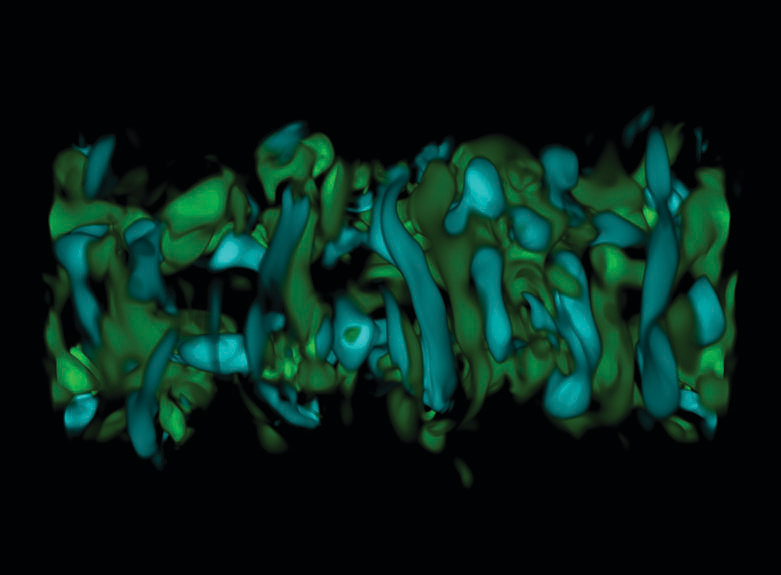}
  \includegraphics[width=0.4\linewidth]{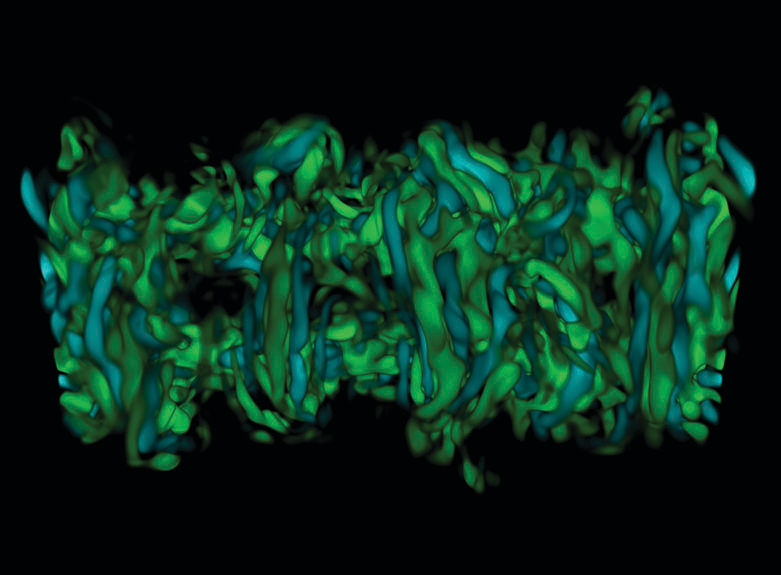}
  \includegraphics[width=0.4\linewidth]{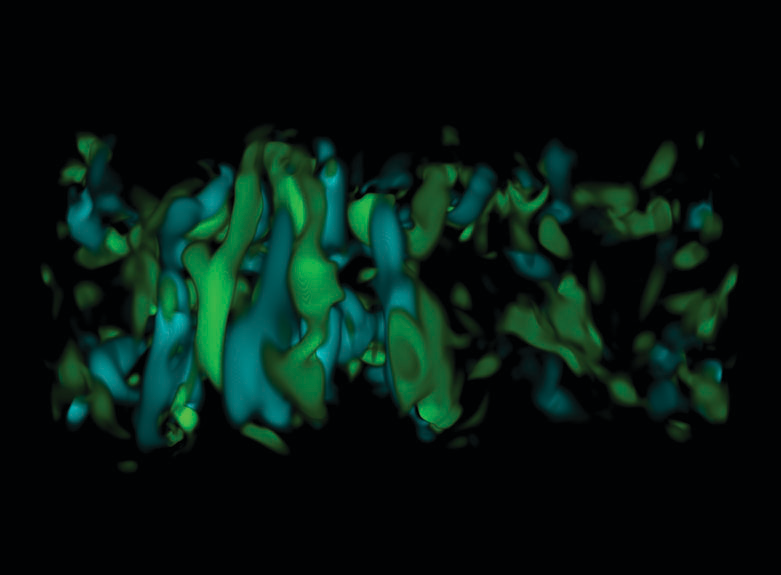}
  \includegraphics[width=0.4\linewidth]{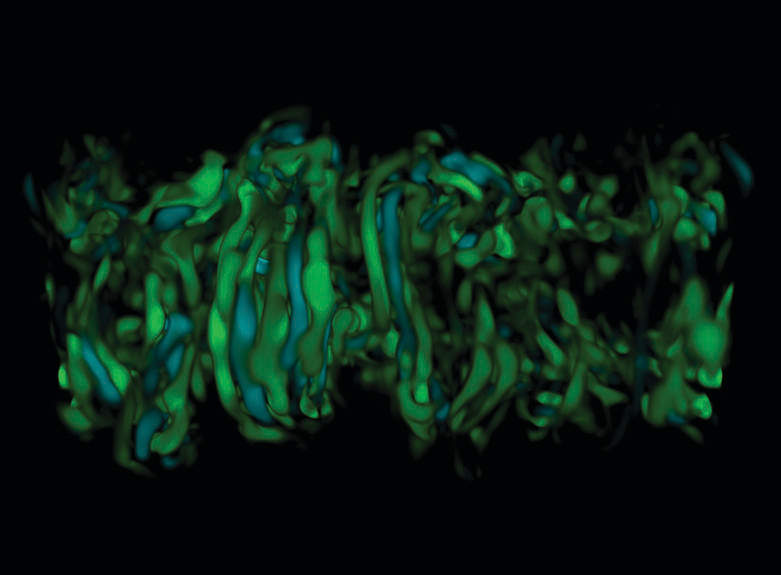}
  \includegraphics[width=0.4\linewidth]{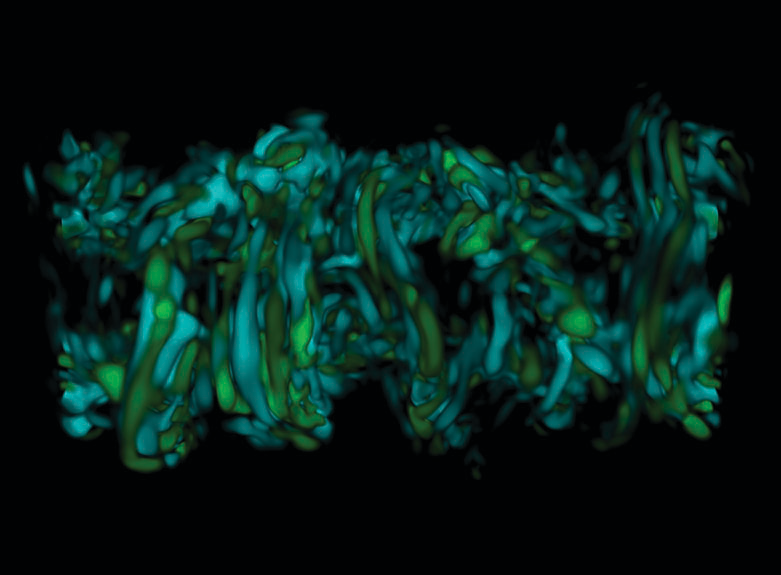}
  \caption{Visualization of the turbulent state reached with two initially antiparallel vortex tubes, when the energy dissipation rate is maximized, at $t=82.3$.
(a) Visualization of the vorticity modulus. (b-g) Localized regions of intense energy transfer for (b) $\mathcal{T}_{1,2}$, (c) $\mathcal{T}_{1,3}$, (d) $\mathcal{T}_{1,4}$, (e) $\mathcal{T}_{2,3}$, (f) $\mathcal{T}_{2,4}$, (g) $\mathcal{T}_{3,4}$. Regions of intense positive energy transfer from larger to smaller scales are blue and regions of intense negative energy transfer from smaller to larger scales are green. 
}
\label{fig:tubes_peak_dissipation_visualization}
\end{figure}

Figure \ref{fig:tubes_spectra}(a) shows how the flow develops the classical $E(k) \propto \epsilon^{2/3} k^{-5/3}$ spectrum, which coincides 
with the peak in the instantaneous dissipation rate~\citep{McKeown:2020}.
It is therefore of fundamental interest to 
study energy transfer in this regime, where we observe a turbulent spectrum in a flow which may be viewed as conceptually simpler than the HIT flow studied in Section~\ref{sec:HIT}.
For future comparison with the HIT configuration,
we note that the Taylor Reynolds number of the flow studied in ~\cite{McKeown:2020,Ostilla:2021} is estimated to be of the order of $Re_\lambda \approx 50$, which indicates that the inertial range encompasses a more restricted range of spatial scales compared to our HIT run at $Re_\lambda = 210$. This means that the $I_5$ shell will be affected significantly by dissipation, and cannot be considered to belong to the integral range. The transfers of energy from $I_5$ to smaller scales will accordingly be weaker than those seen in HIT. 

Fig.~\ref{fig:tkq_tubes_peak_dissipation} shows the energy transfer statistics for the flow at time $t=82.3$, which corresponds to the peak in the energy dissipation rate. 
Fig.~\ref{fig:tkq_tubes_peak_dissipation}(a) shows the normalized integrated 2D energy transfer spectrum, $T_{K,Q}/\epsilon$, at this time, which reveals an ongoing transfer of energy towards small scales, mostly concentrated along the diagonal. The large energy transfer in $T_{1,3}$, clearly visible at earlier times, shown in Fig.~\ref{fig:3}(a), remains present, albeit in a much weaker form, as it
is now smaller than $T_{1,2}$. This is highlighted in Fig.~\ref{fig:tkq_tubes_peak_dissipation}(b), where the scaling of the $T_{K,Q}/\epsilon$ term is shown. Our analysis of the third-order correlator, $T_{K,P,Q}/\epsilon$ (see Eq.~\ref{eq:def_T3_int}) for HIT, shown in Fig.~\ref{fig:tkq_tubes_peak_dissipation}(c-d) reveals similar features. The pattern is qualitatively similar to that of HIT, shown in Fig.~\ref{fig:tkq_hit}(c-d), but the $T_{2,3}$ shell is not as active as in homogeneous isotropic turbulence, due to the memory of the initial conditions of the vortex tubes.

The transient turbulent flow which results from the iterative cascade of instabilities documented in~\cite{McKeown:2020}, exhibits a complicated structure, clearly visible from the structure of the vorticity modulus, shown in Fig.~\ref{fig:tubes_peak_dissipation_visualization}(a).
Furthermore, Fig.~\ref{fig:tubes_peak_dissipation_visualization}(b-g) shows all the energy transfer interactions between the first four shells. Remarkably, at all scales represented, the intense contributions to the energy transfer are observed in the vicinity of the interacting vortex pairs which were formed as a result of the elliptical instability. This is suggestive of the important role played by these vortex pairs and by their iterative interactions in transferring energy across scales. Although it is clear that most intense regions of energy transfer, be it towards small scales or towards large scales (blue or green, respectively) occur very often between pairs of vortex tubes, it is not clear whether a given pair of vortex tubes will give rise to a transfer towards small or towards large scales; both cases can be observed, depending on the particular pair $(K,Q)$ considered. This extends the observations in HIT, Section~\ref{sec:HIT}, that the energy transfer associated with intense structures, can be either up- or down-scale. 

As it was the case at earlier times, see Section~\ref{subsec:par_early}, we can confirm the visual intuition of correlations between the structures of intense regions of $\mathcal{T}_{K,Q}$ and $\omega_Q$ by determining the Pearson correlation coefficients. First, we find a high correlation coefficient of $\approx 0.8$ between $S_Q$ and $\omega_Q$ for the shells $I_2$, $I_3$ and $I_4$, similar to what was reported at earlier times. This indicates that regions of high vorticity and high strain correspond to each other in this flow, unlike in HIT. 

We also find a value of the correlation coefficient of $\approx -0.15$ between $\mathcal{T}_{2,3}$ and $\omega_2$ and $S_2$, as well as with $S_3$ and $\omega_3$. In addition, we find a similar correlation coefficient between $\mathcal{T}_{3,4}$ and all of $\omega_3$, $\omega_4$, $S_3$ and $S_4$. By calculating the correlation coefficient for $|\mathcal{T}_{K,Q}|$ with the strains and vorticities mentioned above, we find that it is consistently of the order $0.5-0.6$ for the pairs $(K,Q)=(2,3)$, $(2,4)$ and $(3,4)$, as was seen at earlier times. This correlation indicates a much closer overlap between regions of high strain and vorticity and those of intense energy transfer than what was observed for HIT (see Section~\ref{sec:HIT}). Furthermore, in this late stage, the presence of a negative correlation between the signed $\mathcal{T}_{K,Q}$ and vorticity and strain shows that there is weak preference for transport to smaller scales associated with intense vorticity or strain.

While the configuration of two originally antiparallel tubes leads to a flow exhibiting a transient $k^{-5/3}$ energy spectrum, along with an integrated energy transfer spectrum reminiscent of HIT flows, it differs in the correlations between strain and vorticity. This flow can still be useful as a conceptually simple configuration for testing and visualizing hypotheses about the role of vortex interactions in the conveyance of energy to smaller scales. However, it should be emphasized that the interaction between two vortex tubes does not fully encompass the multi-scale interactions needed in an accurate model for the energy cascade.
In Appendix \ref{subsec:90}, we additionally consider the distinct case of two vortex tubes originally perpendicular to each other. 
The dynamics in this flow leads to the rapid production of small-scale of motion. For this reason, this configuration has been suggested as a canonical flow to study the turbulent cascade~\citep{Yao:2022}.
As clearly demonstrated by~\cite{Ostilla:2021}, however, the dynamics are completely different from those considered in this section.
Here, the tubes initially stretch and flatten prior to contacting one another, producing very thin vortex sheets. The influence of the fluid viscosity at this scale eventually becomes significant, 
which leads to the reconnection of the vortex tubes. Small-scale vortices are generated away from the region where the vortices reconnect~\citep{Ostilla:2021}. The visualization of the energy transfer and comparison of $T_{K,Q}$ and other statistics to those of HIT, presented in Appendix~\ref{subsec:90}, highlights significant differences with the flows studied in Sections~\ref{sec:HIT} and \ref{sec:t-dependent-Fourier}, so care must be taken when using this flow as a model for the energy cascade.

\section{The role of the large-scale strain and vorticity fields}
\label{sec:Goto}

The discussion in section~\ref{sec:HIT} did not provide any indication that the most intense structures of the vorticity field are particularly important to understand the energy transfer in HIT. The complexity of the interactions between scales in the flow comes, to a large extent, from the multiplicity of scales involved, including at the largest scales. The analysis of the two anti-parallel tubes has shown that with a carefully chosen initial condition, the large-scale flow structures can be ordered, and that this can help us directly visualize this transient energy cascade.

In this section, we consider the
possibility of constraining the largest scales of the flow in a statistically stationary manner, in an attempt to simplify the complex dynamics analyzed in Section \ref{sec:HIT}.
We are motivated by the work of~\cite{Goto:2017} who showed that a relatively simple flow configuration at large scales developed a sustained turbulent state, where the identification of certain features associated with vortex stretching corresponded to the transfer of energy across scales.
Following \cite{Goto:2017}, we study here
a configuration forced by a simple forcing mode, $\mathbf{k} = (1, 1, 0)$ in Fourier space (see Eq.~\ref{eq:GotoF}), which leads to a large scale organization of the flow with four columnar vortices arising in a sustained manner. 
In the statistically stationary flow, the results of \cite{Goto:2017,Yoneda:2021} suggest that 
the resulting vorticity field is organized in a hierarchical manner. Namely, the stretching that amplifies the vorticity in a waveband $I_K$ is induced by vorticity in bands $I_Q$, with a length-scale larger than yet close to $I_K$. This process iterates in a self-similar manner down the scales from the first four large-scale tubes to the dissipation scale. This enticing self-similar picture is reminiscent of the one proposed by~\cite{Tennekes:1972}, although based on a qualitatively different picture of the role of strain in the energy transfer across scales. The scenario proposed by~\cite{Goto:2017} also shares qualitative similarities with the iterative cascade in the configuration of two antiparallel vortex tubes~\citep{McKeown:2020}. However, recent works indicate that the role of vortex stretching in the cascade of energy is subdominant, compared to the amplification of strain \citep{Carbone:2020,Johnson:2020,Johnson:2021}. 

Fig.~\ref{fig:Gotostats} provides an elementary characterization of the turbulence in the flow. The Reynolds number is estimated to be
$Re_\lambda\approx140$. Understandably, the flow deviates from the typical $k^{-5/3}$ spectra at the largest scales, as shown in Fig.~\ref{fig:Gotostats}(a), or the energy flux spectrum $\Pi(k)/\langle\epsilon\rangle$, as shown in Fig.~\ref{fig:Gotostats}(b). However, it is in these largest scales that we can visually appreciate the organizing effect of the forcing. The visualization of $\omega_1$ 
 in Fig.~\ref{fig:Gotostats}(c) showcases the four tubes generated by
the forcing. In $\omega_3$, shown in Fig.~\ref{fig:Gotostats}(d), we observe the hierarchical arrangement suggested by \cite{Goto:2017}, where the vorticity preferentially aligns the next hierarchy of filaments perpendicular to the large-scale vortex tubes. The vorticity filtered at $I_5$ and the unfiltered vorticity, indicated in Fig.~\ref{fig:Gotostats}(e-f) respectively, show less preferential alignment,though the flow is clearly anisotropic. We also note that the size of the most intense structures of the vorticity field, shown in Fig.~\ref{fig:Gotostats}(f), are smaller than those of HIT at $Re_\lambda = 210$, shown in Fig.~\ref{fig:HIT_intro}. This observation could be explained by the reduced Reynolds number of the flow illustrated in Fig.~\ref{fig:Goto_tkq_visualization}, $Re_\lambda \approx 140$, which is less than that in the HIT flow shown in Fig.~\ref{fig:HIT_intro}. 

\begin{figure} 
 \centering
 \subfigure[]{\includegraphics[width=0.45\linewidth]{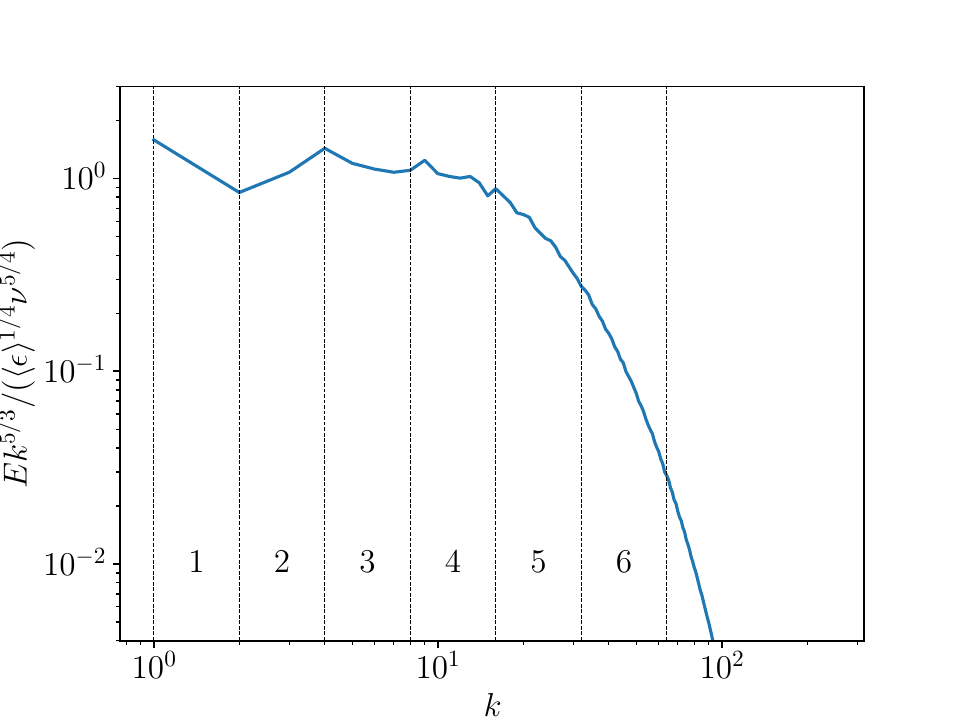}
 }
 \subfigure[]{\includegraphics[width=0.45\linewidth]{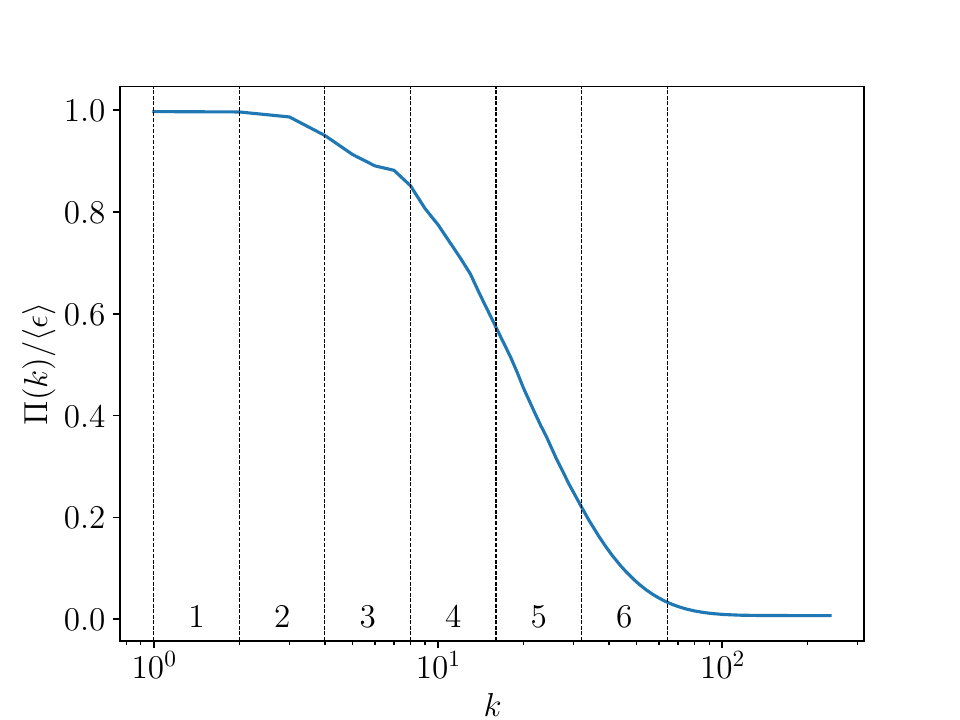}
 }
 \includegraphics[width=0.45\linewidth,trim={220 200 220 180},clip]{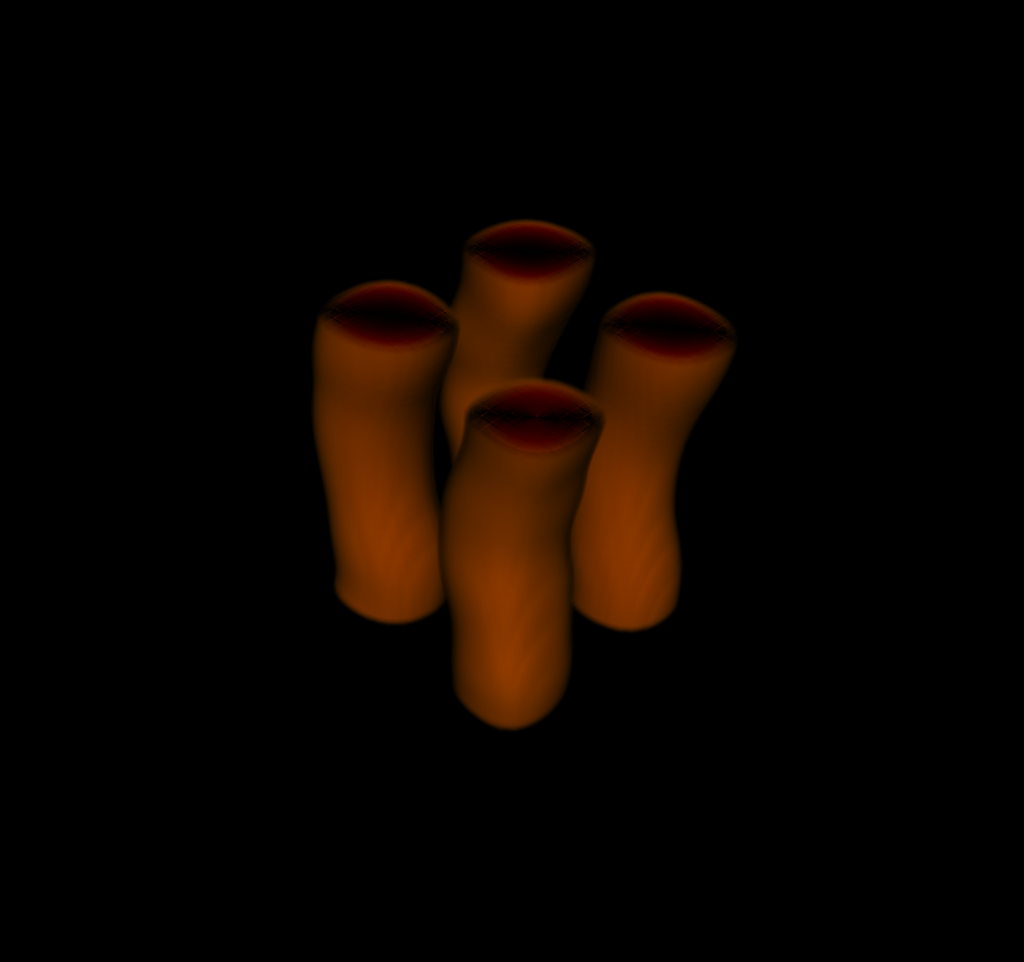}
 \includegraphics[width=0.45\linewidth,trim={220 200 220 180},clip]{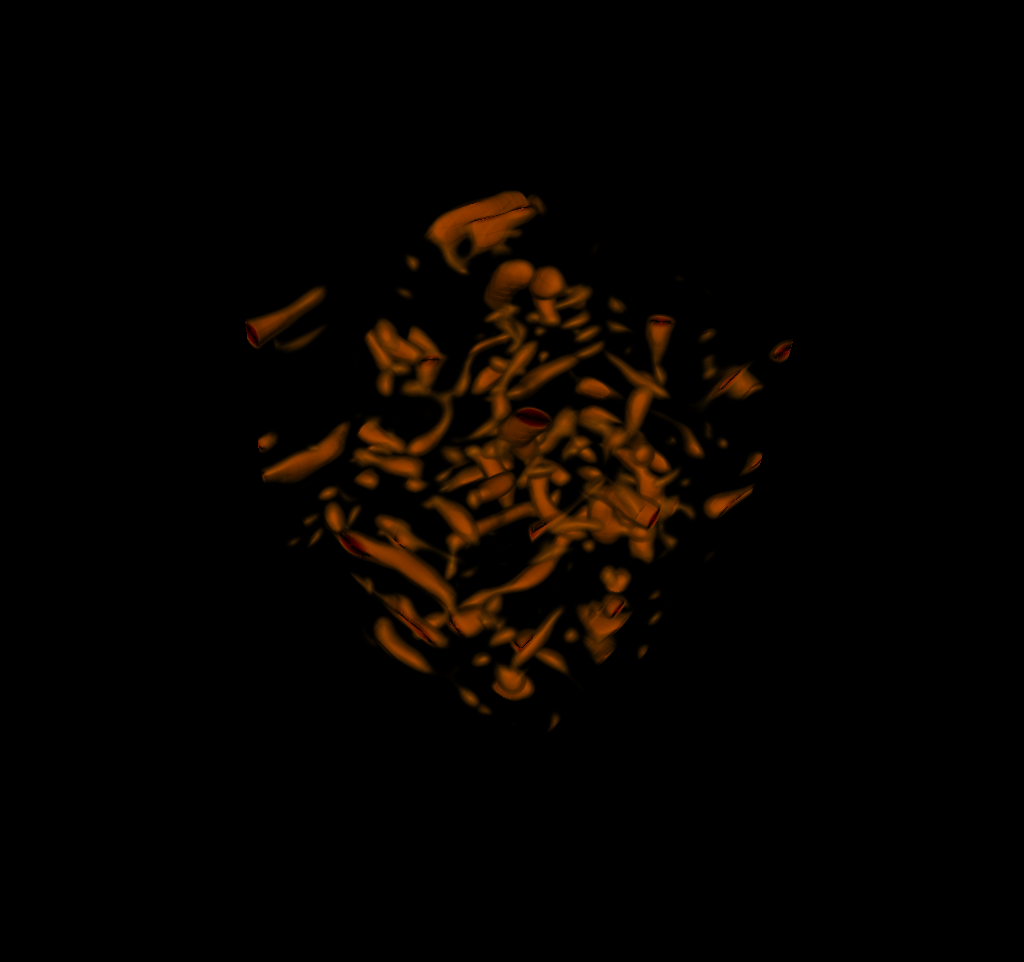}
 \includegraphics[width=0.45\linewidth,trim={220 200 220 180},clip]{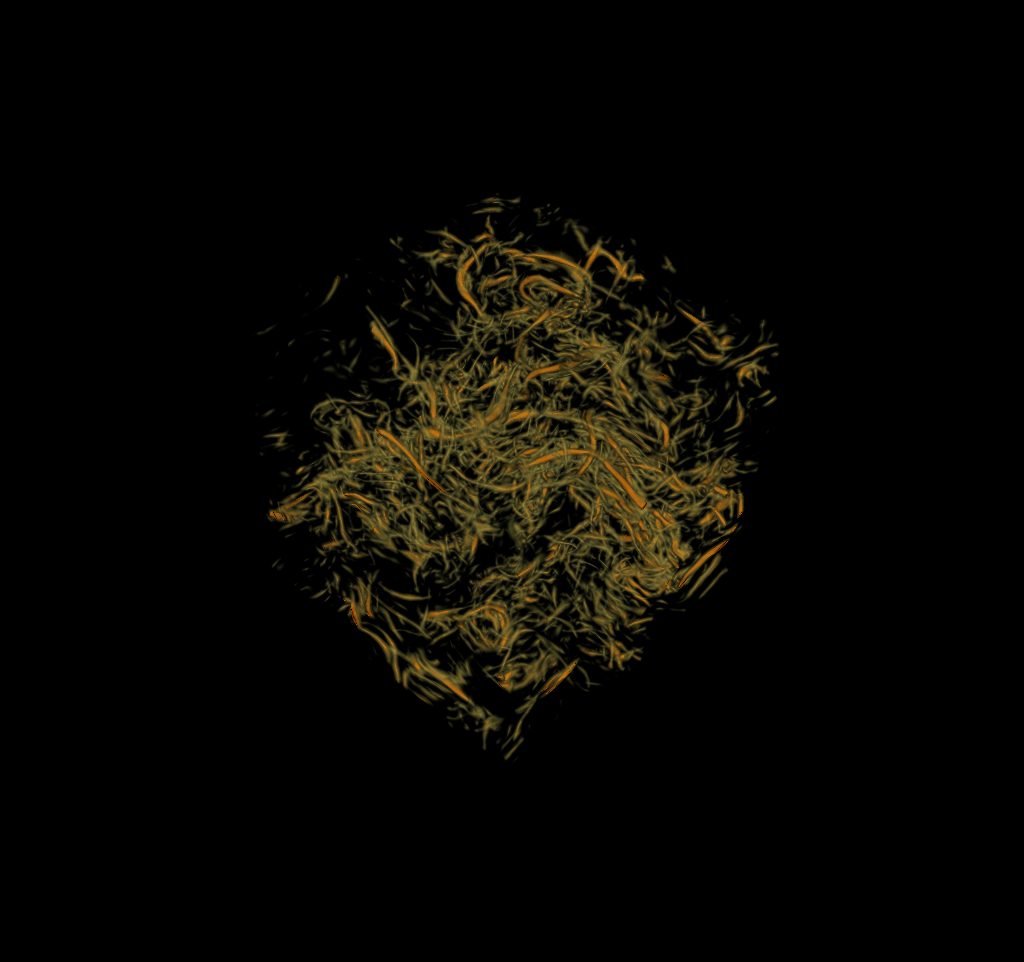}
 \includegraphics[width=0.45\linewidth,trim={220 200 220 180},clip]{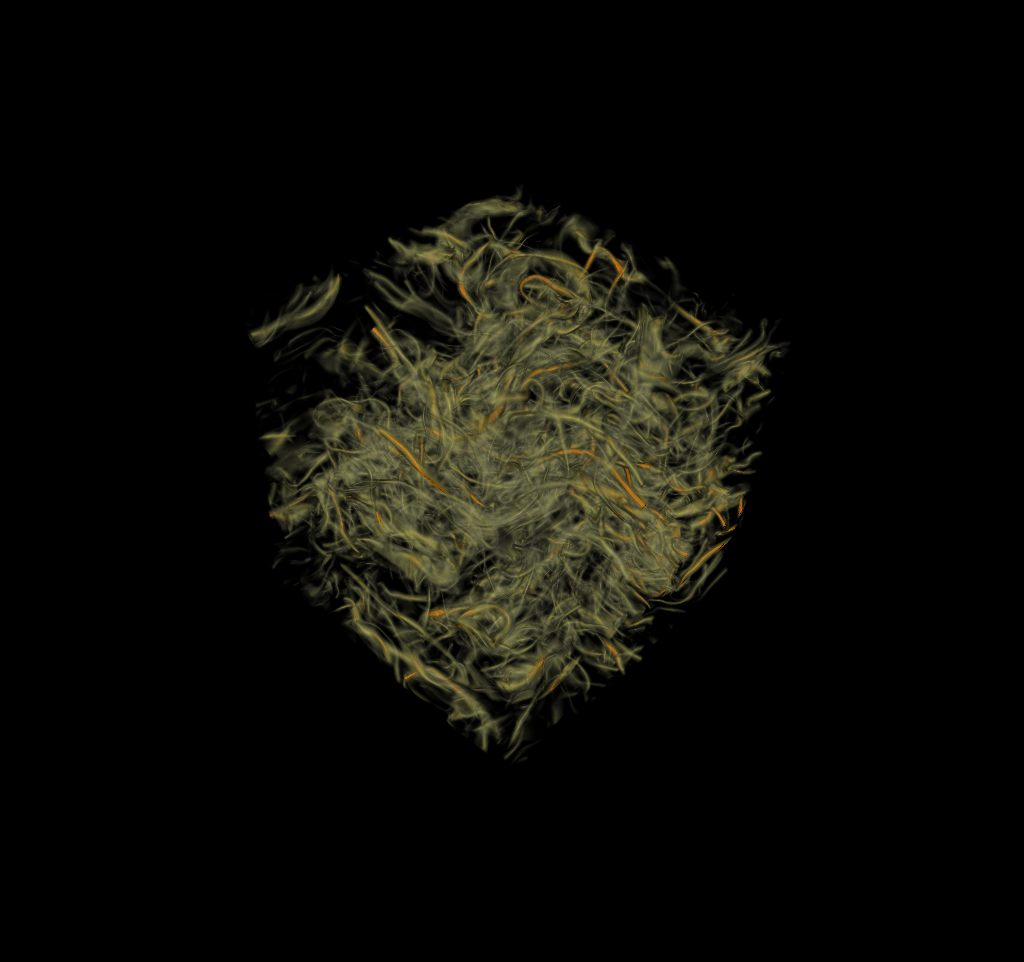}
  \caption{Statistics for DNS with the forcing used in \cite{Goto:2017}. (a) Shows the compensated energy spectrum $Ek^{5/3}/(\langle\epsilon\rangle^{1/4}\nu^{5/4})$, where $\epsilon$ is the dissipation rate and $\nu$ is the kinematic viscosity of the fluid. (b) Shows the energy flux spectrum $\Pi(k)/\langle \epsilon \rangle$. (c-f) Show the vorticity filtered at scales $\omega_1$, $\omega_3$, $\omega_5$, and the unfiltered vorticity $\omega$, respectively, where regions of low vorticity have been made transparent. }
\label{fig:Gotostats}
\end{figure}

To further study this flow, in Fig.~\ref{fig:tkq_Goto} we show an analysis of energy transfer at a time after the flow has reached a statistically steady state. Fig.~\ref{fig:tkq_Goto}(a) shows the integrated 2D energy transfer spectrum, $T_{K,Q}$. Beyond a qualitative resemblance with comparable figures previously examined, we notice that the transfer from $I_1$ to modes at higher wavenumbers does not decay with increasing wavenumber as fast as in the case of HIT, as shown in Fig.~\ref{fig:tkq_hit}(a). Rather the transfer from the $I_1$ shell extends appreciably to bands beyond $I_3$. This is confirmed in Fig.~\ref{fig:tkq_Goto}(b), which shows a weak decay of $T_{1,Q}$ with increasing $Q$, far from the power laws obtained for HIT in Fig.~\ref{fig:tkq_hit}(b). This is reminiscent of the antiparallel tube interaction, see 
Fig.~\ref{fig:tkq_tubes_peak_dissipation}(a-b). 
For both the anti-parallel tubes and the present flow configuration, the organized nature of the flow in the band $I_1$ leads to an efficient transfer of energy, reaching out to scales (values of $Q$) beyond what would occur in the case of an HIT flow with a much less structured forcing at the largest scales.

Conversely, the transfer $T_{2,Q}$ in Fig.~\ref{fig:tkq_Goto}(b) decays more rapidly than in the HIT case, see Fig.~\ref{fig:tkq_hit}. In addition, the mode $I_2$ seems to transfer very little energy to $I_3$, and this ``step'' of the cascade seems missing, as shown in Fig.~\ref{fig:tkq_Goto}(a). This is further confirmed by examining the triple correlator between 3 bands $T_{K,P,Q}$, as defined by Eq.~\eqref{eq:def_T3_int}, and fixing $P=2$, as shown in Fig.~\ref{fig:tkq_Goto}(c). This shell is absent from the overall transport of energy, unlike for the case of HIT. It thus appears that, although the forcing introduced by \cite{Goto:2017} induces a clear ordering of the velocity at large scale, it does not provide insight to the subtleties of the transport in the large-wavelength shells. Meanwhile, the triple correlator with $P=4$ shell, shown in Fig.~\ref{fig:tkq_Goto}(d), seems largely similar to the case of HIT, indicating that once the strain field becomes largely isotropic, the phenomenology begins to better resemble that of the general turbulent cascade.

\begin{figure}
 \centering
 \subfigure[]{ 
 \includegraphics[width=0.45\linewidth]{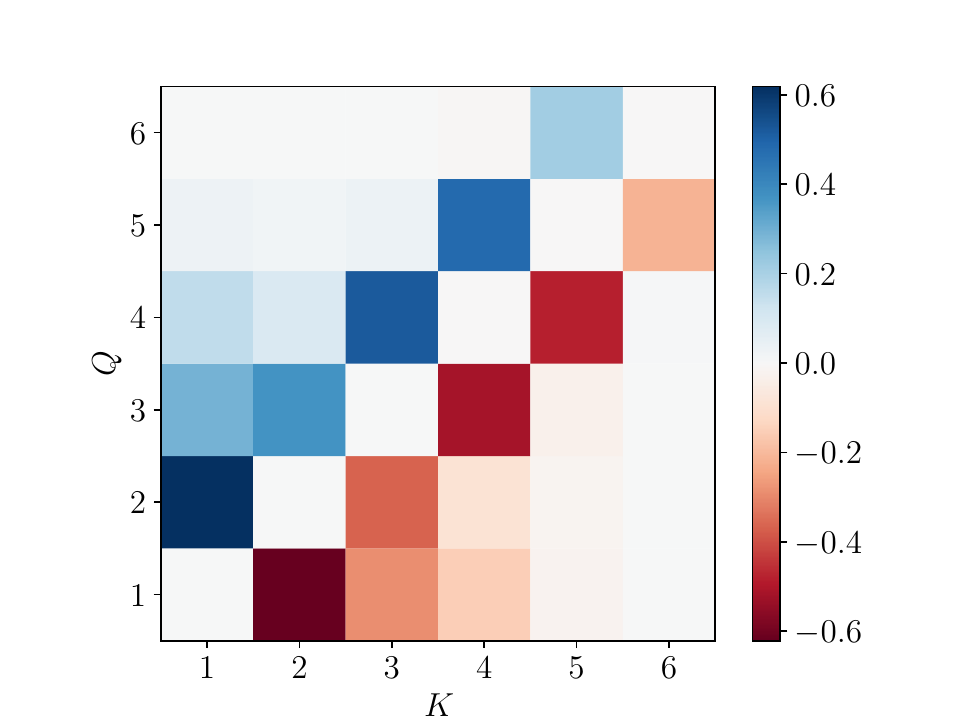}
 }
 \subfigure[]{
 \includegraphics[width=0.45\linewidth]{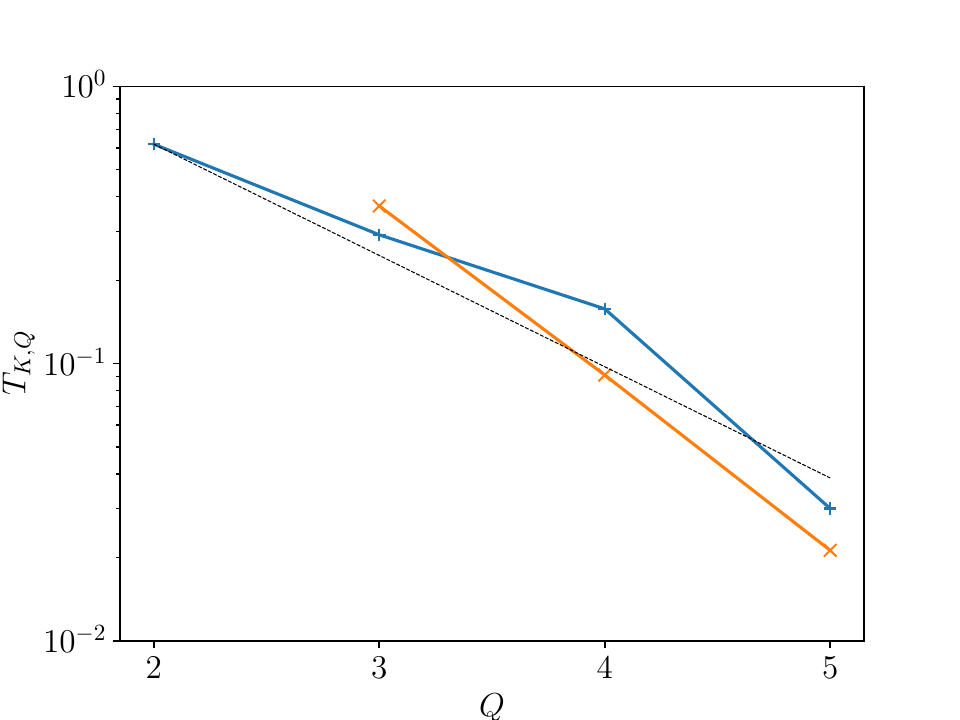}
 }
 \subfigure[]{
 \includegraphics[width=0.45\linewidth]{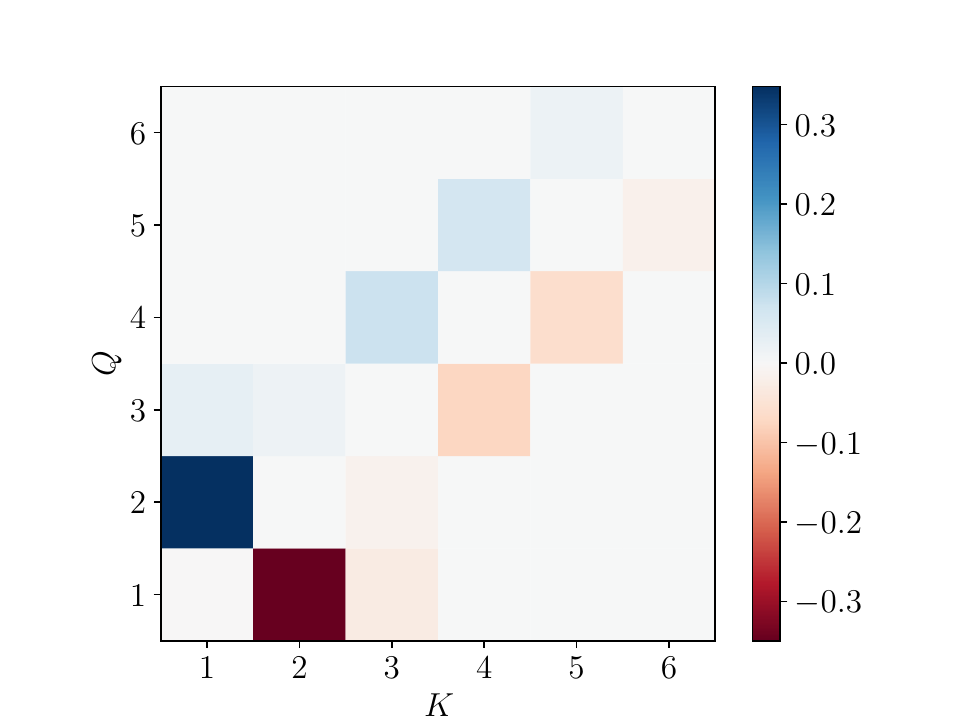}
 }
 \subfigure[]{
 \includegraphics[width=0.45\linewidth]{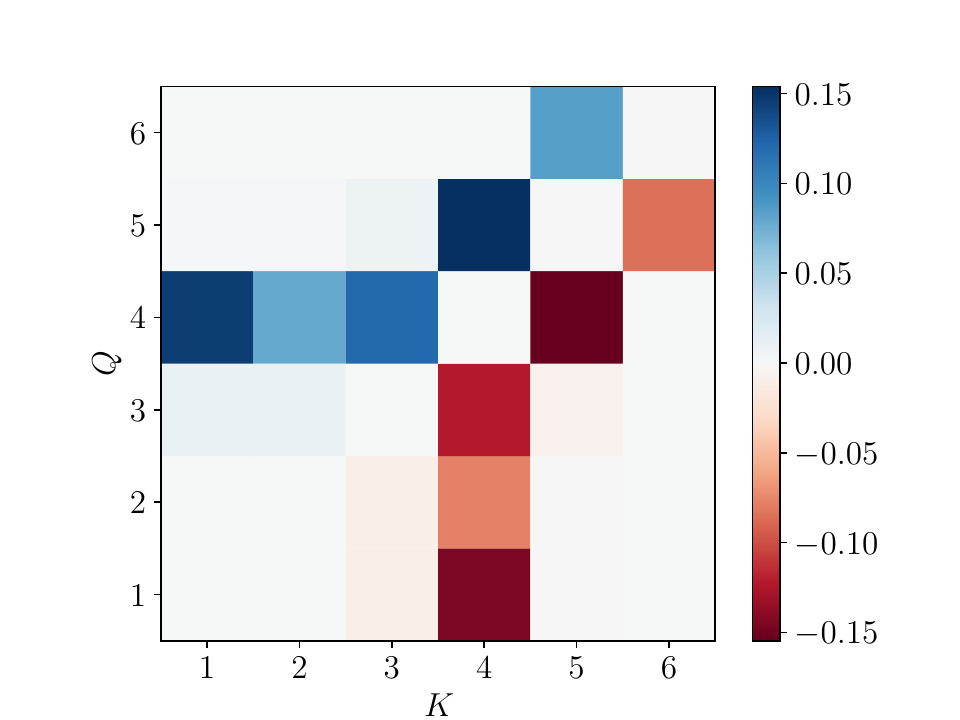}
 }
  \caption{Instantaneous energy transfer 
  Instantaneous integrated 2D energy transfer spectrum, $T_{K,Q}/\langle \epsilon \rangle$ for DNS with the forcing used in \cite{Goto:2017} (a)
at $Re_{\lambda} \approx 140$, where $k_f = 2.3$. The transfer is mostly localized to neighboring bands.
(b) shows the integrated transfer rate, $T_{K,Q}/\langle \epsilon \rangle$  from modes $K = 1$ (blue plus symbols) and from $K = 2$ (red cross symbols). The straight dashed line corresponds to an exponential decay: $\propto 2^{-4 Q/3}$.  The bottom panels show the triple correlator $T_{K,P,Q}/\langle\epsilon \rangle$ for (c) $P=2$ and (d) $P=4$.}
\label{fig:tkq_Goto}
\end{figure}

Fig.~\ref{fig:Goto_tkq_visualization} visualizes the energy transfer rate, $\mathcal{T}_{K,Q}$, in real space. As was the case in HIT (see Fig.~\ref{fig:tkq_vis_hit}(a-f), Fig.~\ref{fig:Goto_tkq_visualization} shows that regions of intense energy transfer from $I_K$ to $I_Q$ with $Q>K$ (blue regions), and from $K > Q$ (green regions) are somewhat spread throughout the flow, without any strong correlation with the most intense structures of the vorticity field. Our attempts to correlate the energy transfer rate $\mathcal{T}_{1,2}$ with the regions where $\omega_1$ did not reveal as strong a correlation as the one visually observed by \cite{Goto:2017} when analyzing vortex stretching. More generally, as was the case in the HIT configuration (see Fig.~\ref{fig:tkq_vis_hit}(a-f)), we did not observe any convincing correlation between regions where $\mathcal{T}_{K,Q}$ is intense and regions of intense values of $\omega_K$ or $\omega_Q$.

\begin{figure}
\centering
 \includegraphics[width=0.45\linewidth]{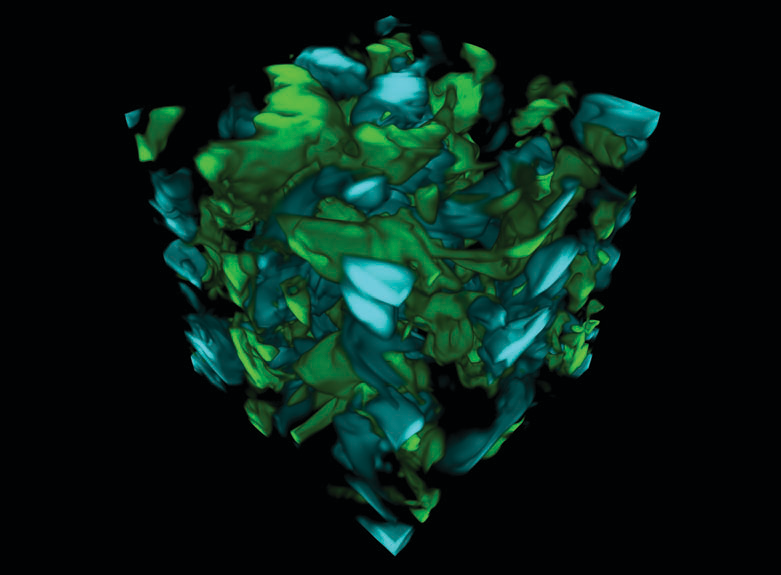}
 \includegraphics[width=0.45\linewidth]{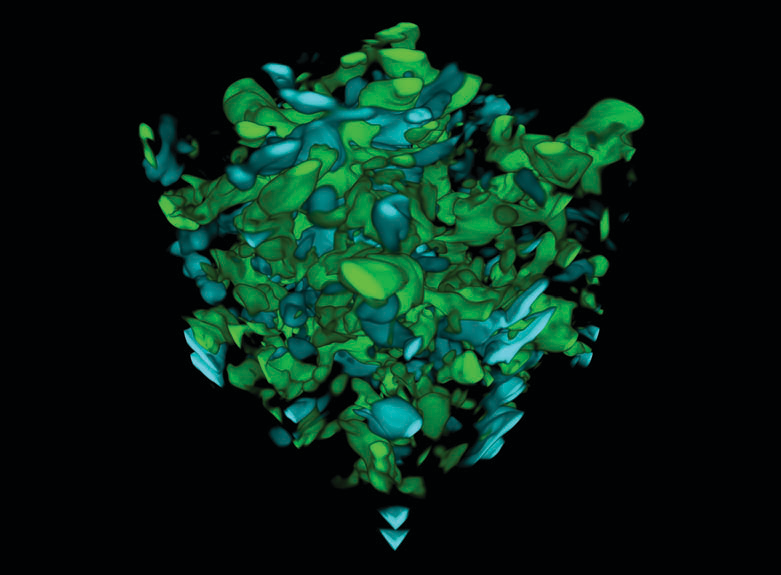}
 \includegraphics[width=0.45\linewidth]{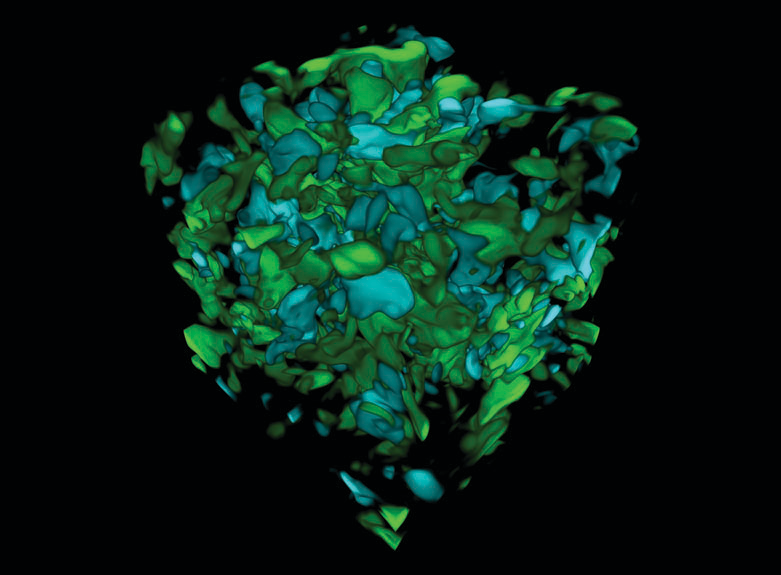}
 \includegraphics[width=0.45\linewidth]{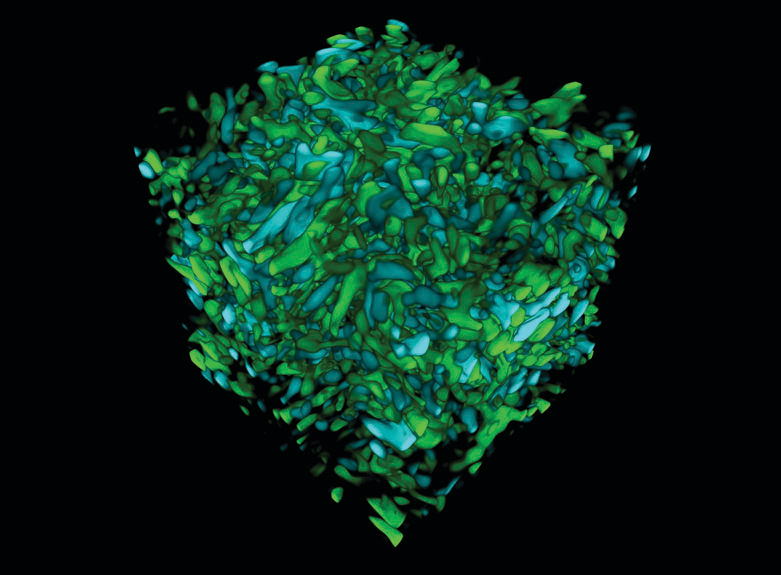}
  \caption{Visualization of the energy transfer rate, $\mathcal{T}_{K,Q}$ for DNS with the forcing used in \cite{Goto:2017}. (a) $\mathcal{T}_{1,2}$, (b) $\mathcal{T}_{1,3}$, (c) $\mathcal{T}_{2,3}$, (d) $\mathcal{T}_{3,4}$. Regions of intense positive energy transfer from larger to smaller scales are blue and regions of intense negative energy transfer from smaller to larger scales are green. 
}
\label{fig:Goto_tkq_visualization}
\end{figure}

The lack of any visual correlation in the energy transfer between scales, $\mathcal{T}_{K,Q}(\textbf{x})$, and the various components of vorticity, $\omega$, and strain, $S$, at a scale $Q$ or $K$, is confirmed by calculating the Pearson correlation coefficient between these quantities. As was the case for HIT considered in Section~\ref{sec:HIT}, we find negligible correlations between $\mathcal{T}_{K,Q}$ and $S_Q$, $S_K$, $\omega_K$ and $\omega_Q$ (on the order of $10^{-2}$). The correlation between $| \mathcal{T}_{K,Q}|$ and strain and vorticity filtered at scales $K$ or $Q$ are slightly larger, but remain small, $\lesssim 0.3$ when compared to those in Section \ref{subsec:par_late}. This confirms the visual impression of a lack of correlation between energy transfer and strain or vorticity at a given scale. In addition, our visualizations and the calculation of the Pearson correlation coefficient do not reveal any particularly strong correlation between strain and vorticity, filtered at the same scale, in agreement with the HIT configuration detailed in Section~\ref{sec:HIT}. 

This leads us to the conclusion that the organized structure of vortex stretching found by~\cite{Goto:2017}, and analyzed by~\cite{Yoneda:2021}, does not particularly help in identifying a simple underlying structure of energy transfer.
This is particularly true at the smallest scales, where the transfer of energy strongly resembles that of HIT.
In this sense, the remarkably simple picture, observed in the interaction of antiparallel vortex tubes and in the configuration studied by~\cite{Goto:2017}, appears to be the result of a strong influence of the large-scale forcing (or initial conditions), even if the flow with two antiparallel tubes shows statistics closer to those of HIT. This makes both flows ideal for the study of the energy transfer associated to vortex stretching, but less applicable to the general case of the turbulent cascade. 

\section{Summary and conclusion}
\label{sec:concl}

In this work, we investigated the transfer of energy between bands of wavenumbers in several previously studied flow configurations in order to study the turbulent energy cascade. To this end, we decomposed the velocity fields by splitting the wavenumber space into shells of wavenumbers, $I_P$, centered around $k_f \times 2^{P-1}$, as defined by Eq.~\ref{eq:I_P}. The energy transfer rate $T_{K,Q}$ between bands of wavenumbers $I_Q$ and $I_K$ is then written as an integral over a scalar quantity, $\mathcal{T}_{K,Q}(\mathbf{x},t)$. This allowed us to characterize not only the energy transfer in Fourier space, by representing $T_{K,Q}$ as a function of $K$ and $Q$, but also to visualize this quantity in real space, characterizing the regions where $\mathcal{T}_{K,Q}(\mathbf{x},t)$ is large. We also examined the possibility of correlating this field with the vorticity and strain, filtered in the same bands of wavenumbers $I_K$ and $I_Q$, to examine how energy transfer coincides with regions of high strain or vorticity.

We first used this decomposition to consider the canonical configuration of a statistically stationary homogeneous isotropic turbulent flow (in the presence of an external forcing), at a moderately high Reynolds number, $Re_\lambda = 210$. In agreement with earlier studies, we verified that in Fourier space, the energy transfer is mostly local, i.e. $T_{K,Q}$ decreases precipitously with $|K-Q|$. Our numerical results are consistent with earlier numerical works~\citep{Alexakis:2005,Mininni:2005}, which found that $T_{K,Q} \sim |K-Q|^{-4/3}$, in agreement with earlier predictions of~\cite{Kraichnan:1966,Tennekes:1972,Eyink:2009}.

While the structure of the energy transfer for HIT in Fourier space, as defined by the integrated second-order correlator $T_{K,Q}$, has a remarkably simple structure, we could not find any clear structure of the energy transfer function, $\mathcal{T}_{K,Q}$ in real space. Visually, this function neither correlates with the regions of most intense vorticity or strain in the flow, nor with the vorticity or strain filtered in bands $I_K$ or $I_Q$, as defined by Eq.~\ref{eq:I_P}. This lack of correlation was validated by by calculating the Pearson correlation coefficient between these sets of variables. We also observed that the vorticity and strain, themselves, are weakly correlated in real space. We notice that the prediction of~\cite{Tennekes:1972} about the locality of the energy transfer was based on a physical picture grounded on an analysis of the effect of strain at the largest scale. The lack of correlation between $\mathcal{T}_{K,Q}(\mathbf{x})$ and the strain $S_Q$ thus appears unexpected in this respect. 

We then considered the transfer of energy in time-dependent flows, taking as initial condition two laminar, antiparallel counter-rotating vortices, which rapidly break down and develop intense small-scale motions and eventually generate  a transient $k^{-5/3}$ energy spectrum. Using our method, we co-localized the intense vortex structures, which develop in the flow as a result of the interaction between the two original vortices, with the regions of intense energy transfer directed mostly towards small scales. This simple correlation between intense vortices and regions of intense energy transfer persists even in the turbulent regime, with a $k^{-5/3}$ energy spectrum. We noticed, however, that the correlation between the energy transfer function, $\mathcal{T}_{K,Q}(\mathbf{x)}$ and vorticity or strain, filtered at similar scales, remained relatively small, even if larger than for the HIT case. This implies that in these seemingly simple flows, the interaction between scales 
is much more complicated than what one could have inferred from the visualisations of the flow. 

We finally attempted to use this method on the forcing proposed by \cite{Goto:2017}, which structures the large scale-structures while resulting in a statistically stationary turbulent flow. In their original studies, this forcing was very useful in clarifying the structure of vortex stretching in the flow. However, visualizing $\mathcal{T}_{K,Q}$ did not particularly aid in clarifying the structure of energy transfer in the flow.
It is thus interesting to contrast the results shown here with the simpler picture obtained by analyzing vortex stretching by \cite{Goto:2017} and \cite{Motoori:2019}. Vortex stretching reveals a more intuitive picture, whereby the strain generated at scale $I_K$ correlates strongly with vorticity, filtered at a scale $I_{K + 1}$ or $I_{K+2}$. This picture emerges clearly from the analysis of the flow considered in Section~\ref{sec:Goto}, for $K=1$ and $Q=3$, and is qualitatively similar to what has been obtained with two parallel tubes by~\cite{McKeown:2018,McKeown:2020}.  
It is important to recall, however, that the integrated energy transfer rate, $T_{K,Q}$ (and the corresponding scalar function, $\mathcal{T}_{K,Q}(\mathbf{x})$ ), fundamentally differs from the observables used to characterize vortex stretching by \cite{Goto:2017} and \cite{Yoneda:2021}. 
As stressed recently~\citep{Carbone:2020,Johnson:2021}, vortex stretching does not coincide with energy transfer. In fact, nonlinear interactions involving strain contribute more to energy transfer than vortex stretching, except at the smallest, dissipative scales of the flow. Hence, what is revealed by examining vorticity amplification appears unlikely to be related to energy transfer. Indeed, even in configurations where the analysis of \cite{Goto:2017} provides a clear scaling argument of vortex stretching, as studied by \cite{Yoneda:2021}, we could not identify any particularly strong correlation between energy transfer with intense vorticity nor with intense strain at the various scales considered. 
 
The additional flow configuration of two initially perpendicular vortex tubes (see Appendix \ref{subsec:90}), which leads to reconnection via a strong flattening of the cores followed by their subsequent breakdown, has also been often considered  an ideal testing ground to study the energy cascade. Although a turbulent regime has been observed at Reynolds numbers higher than what has been studied here, our results reveal a correlation between vorticity, strain, and energy transfer at similar scales that is strongest  in the regime where the two tubes are about to reconnect. This regime, however, does not appear to accurately capture the transfer of energy that occurs in HIT.



Although our work encompasses a limited range of Reynolds numbers ($R_\lambda \le 210$), we do not expect that increasing $R_\lambda$ will lead to any significant change in the complex picture discussed, based on the  numerical data used in this study. 

\acknowledgments{We acknowledge the Research Computing Data Core, RCDC, at the University of Houston for providing us with computational resources and technical support.
AP was supported by the International Research Project ‘Non-Equilibrium Physics of Complex Systems’ (IRP-PhyComSys, France-Israel), and by the French Agence Nationale de la Recherche, under Contract No. ANR-20-CE30-0035 (project TILT). MPB acknowledges the Simons Foundation and ONR N00014-17-1-3029.}

\emph{Declaration of interests:} The authors report no conflicts of interest.

\bibliographystyle{jfm}

\bibliography{Energy_Transfer}

\appendix

\section{Initially perpendicular vortex tubes: reconnection and energy transfer}
\label{subsec:90}

Here, we consider the flow configuration of two initially perpendicular vortex tubes and examine how their interaction and breakdown relates to the energy transfer behaviors detailed in Sections \ref{sec:HIT} and \ref{sec:t-dependent-Fourier}. The interaction between these tubes generates small-scale flow structures, but not the classic $k^{-5/3}$ spectra at the Reynolds number in our simulations, as shown in Fig.~\ref{fig:tubes_spectra_appendix}(a). In addition, we also show the energy flux $\Pi(k)$ in  Fig.~\ref{fig:tubes_spectra_appendix}(b). While energy is transferred to smaller scales, this transfer is much smaller than for the anti-parallel tubes analyzed earlier.

\begin{figure}
 \centering
 \subfigure[]{
 \includegraphics[width=0.48\linewidth]{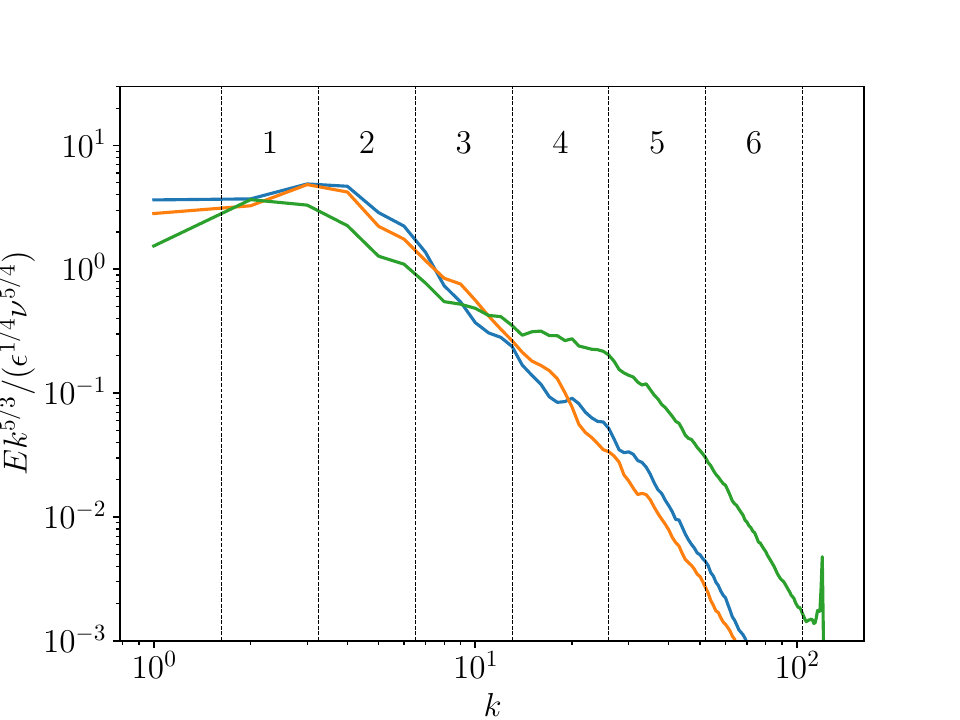}
 }
 \subfigure[]{
  \includegraphics[width=0.48\linewidth]{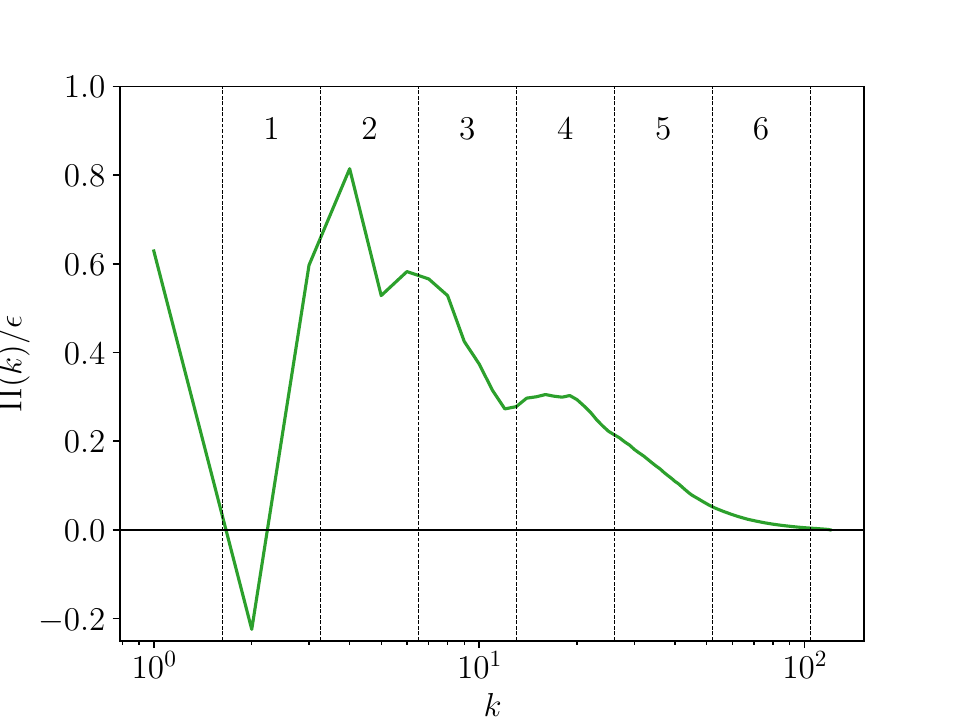}
  }
  \caption{Instantaneous compensated energy spectra (a)  for initially perpendicular tubes at times $t=13$, $18.5$ and $28.5$ and (b) energy flux spectrum $\Pi(k)/\epsilon$, normalized by the dissipation rate, $\epsilon$, at the $t=28.5$. }
\label{fig:tubes_spectra_appendix}
\end{figure}

In this configuration, the dynamics lead to the formation of intense vortex sheets as the tubes initially stretch and flatten prior to contacting one another. These vortex sheets then burst, leading to a reconnection of the vortex tubes which generates a residual a cloud of fine-scale vortices~\citep{Ostilla:2021}. This evolution shares many features with the more classical configuration used to study vortex reconnection that starts with two initially close, antiparallel vortices with a perturbation in their position ~\citep{Brenner:2016,Yao:2022}. The run analyzed here is described in detail in \cite{Ostilla:2021}. The Reynolds number of this run, $Re_\Gamma = 4000$, is sufficient to replicate most of the phenomena observed in symmetric configurations studied at higher Reynolds numbers, e.g. by \cite{Yao:2020}. However, it is still too low to observe evidence of a genuinely turbulent flow with a characteristic $k^{-5/3}$ energy spectrum. 

Fig.~\ref{fig:perpendicular_tubes}(a) shows the integrated energy transfer spectrum, $T_{K,Q}$ at 3 different stages of the evolution: as the vortices approach each other ($t=13$, left column), shortly before they reconnect ($t=18.5$, center column), and after reconnection ($t=28.5$, right column), while the vorticity field at these times is shown in Fig.~\ref{fig:perpendicular_tubes}(b). The maximum kinetic energy dissipation rate, defined as $\epsilon = \nu \langle \omega^2 \rangle$, where $\langle . \rangle$ represents an average in space and time, grows by $\sim 50 \%$ shortly after the time shown in the central column, in the time range $20 \lesssim t \lesssim 24$. 

Even at the earliest time shown, significant energy transfer occurs
between shells $I_1$ and $I_2$, and between shells $I_2$ and $I_3$,
while energy transfer from higher modes at shells ($K \ge 3$) to $Q > K$ is much weaker. When the strong interaction between the tubes commences just prior to reconnection, at $t = 18.5$,  the transfer of energy between $I_1$ and $I_2$ decreases significantly and actually changes sign, indicating a net transfer toward the larger scale. Most of the energy transfer to smaller scales originates from the $I_2$ mode, see also Fig.~2 of~\cite{Ostilla:2021}. 

\begin{figure}
\centering
 \includegraphics[width=0.3\linewidth]{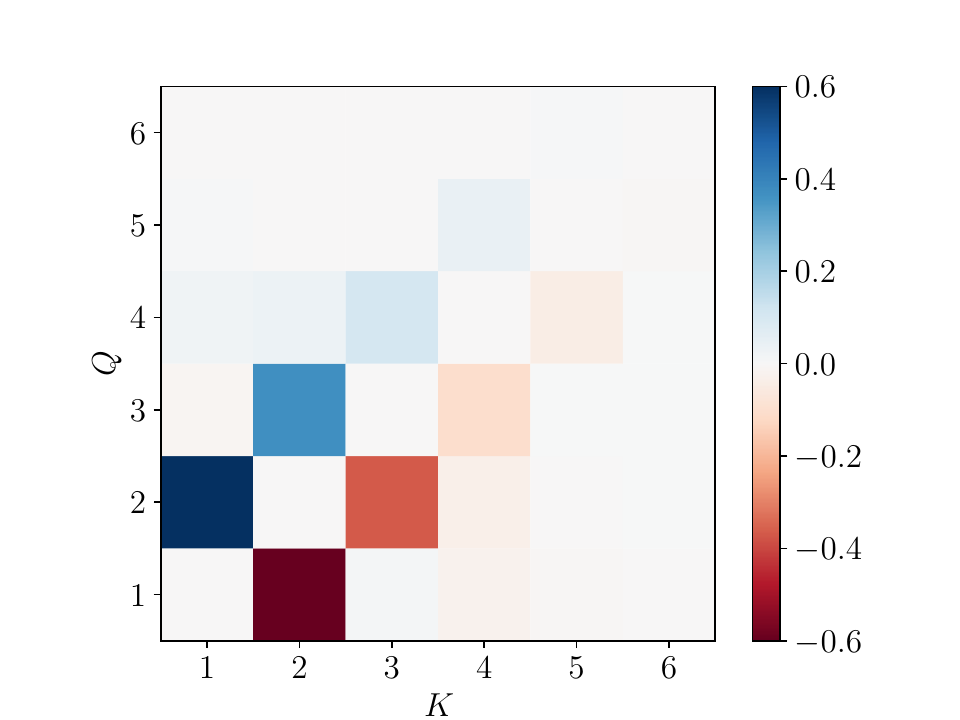}
 \includegraphics[width=0.3\linewidth]{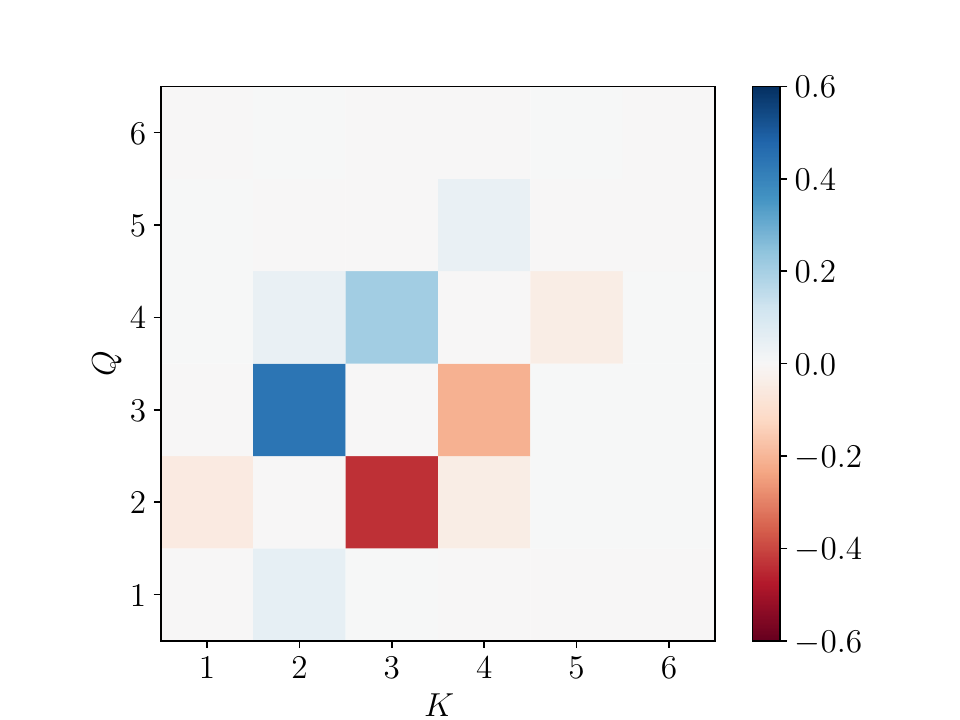}
 \includegraphics[width=0.3\linewidth]{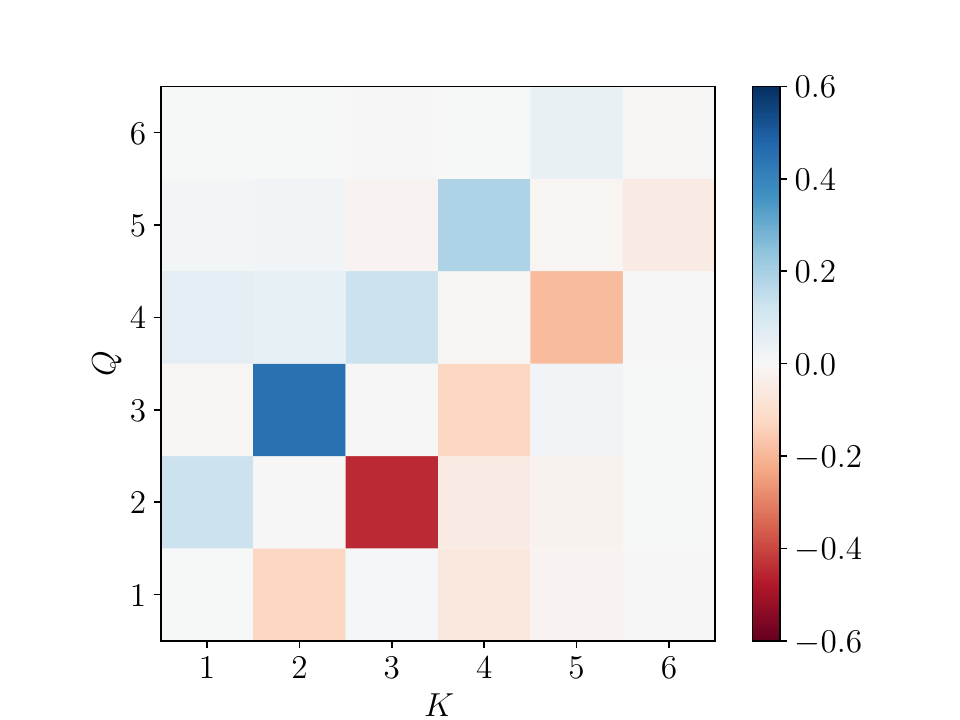}\\
 \includegraphics[width=0.9\linewidth, trim={0 0 0 4.5cm},clip]{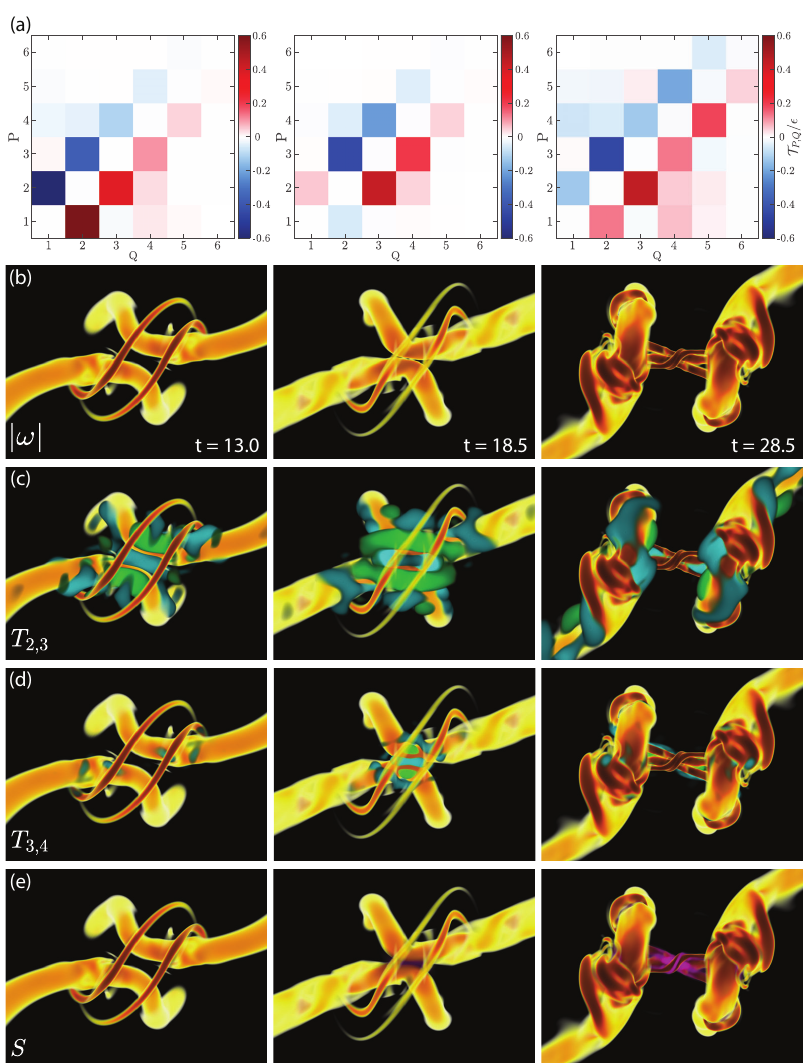} 
\caption{Energy transfer during the reconnection of two initially perpendicular vortex tubes. The flow is shown at three times: 
$t = 13$ as the tubes are approaching (left column); $t = 18.5$ as the tubes begin to interact
(central column), and $t = 28.5$, 
after reconnection (right column). 
(a) 2D energy integrated energy transfer spectrum, $T_{K,Q}/\epsilon$, where $\epsilon$ is the instantaneous energy dissipation rate. The energy transfer is initially confined
to low-wavenumber bands ($K, Q \le 3$), and extends to higher wavenumber-bands
as vortices reconnect.
(b) Front view of the evolution of the vorticity modulus, showing 
the tubes as they approach, interact, and  separate after reconnection, leaving behind a bow structure with intense small-scale motion.
(c) Regions of intense positive energy transfer to smaller scales,
from $I_2 $ to $I_3$ (blue), and of intense negative energy transfer (to 
larger scales) from $I_3$ to $I_2$ (green). Prior to reconnection, intense
energy transfer toward small scales occurs between the two tubes, while some
energy transfer toward larger occurs scales on the outside of the tubes.
Following reconnection, at $t = 28.5$, energy transfer to small scales  
occurrs on the separating tubes.
(d) Regions of intense positive energy transfer to smaller scales from $I_3$
to $I_4$ (blue) and of intense negative energy transfer from $I_4$
to $I_3$ (green). Energy transfer at these smaller scales is concentrated in the reconnection zone
and persist mostly on the bow regions after reconnection. 
(e) Unfiltered rate of strain magnitude (magenta), which is most intense in the spaces between interacting vortices. }
\label{fig:perpendicular_tubes}
\end{figure}

As the energy transfer between wavenumber bands is mostly concentrated between $I_2$ and $I_3$ as well as $I_3$ and $I_4$, we superimpose on the vorticity field  the regions of intense energy transfer $\mathcal{T}_{2,3}$ in Fig.~\ref{fig:perpendicular_tubes}(c) and $\mathcal{T}_{3,4}$ for \ref{fig:perpendicular_tubes}(d). The regions where the energy transfers toward smaller scales (i.e.  $\mathcal{T}_{K,Q}>0$) are shown in blue, and the regions of intense energy transfer toward larger scales (i.e.  $\mathcal{T}_{K,Q}<0$) are shown in green. At the earliest time shown, much of the energy transfer from $I_2$ towards $I_3$ is concentrated in the region where the tubes come closest together, visible as the blue region between the filaments on the left column of Fig.~\ref{fig:perpendicular_tubes}(c). Conversely, we observe a negative transfer toward the larger scale on the other sides of the tubes. At this early time, we do not observe any significant transfer between shells $I_3$ and $I_4$, as shown in Fig.~\ref{fig:perpendicular_tubes}(d).

At $t = 18.5$, when the vortex tubes have begin reconnecting, the band structure of regions of positive and negative energy transfer between $I_2$ and $I_3$ persists, as shown in the middle column of Fig.~\ref{fig:perpendicular_tubes}(c). The positive transfer of energy between the tubes, however, is no longer as strong as it was at $t = 13$. The negative transfer, seen on the other side of the tubes, become simplified. The intense interaction between the tubes gives rise to a strong transfer of energy between the $I_3$ and $I_4$ shells, as shown in the center column of Fig.~\ref{fig:perpendicular_tubes}(d). 

At the later time, $t = 28.5$, after the vortex tubes locally reconnect and separate, Fig.~\ref{fig:perpendicular_tubes}(a) reveals that an active transfer of energy evolves over a broader range of wavenumber bands, and not just around $I_2$, though $T_{2,3}$ is still the largest transfer. At this time, the flow exhibits a characteristic bow-like structure behind the reconnected vortices, with active small-scale motion~\citep{Ostilla:2021}.
The right  column of Fig.~\ref{fig:perpendicular_tubes}(c) shows that regions of positive and negative energy transfer between the $I_2$ and $I_3$ shells are located along the two separating tubes, particularly in the regions where the interacting tubes had reconnected. Also, a significant amount of energy transfer at smaller scales is localized in the two residual braided vortices, as shown in the right column of Fig.~\ref{fig:perpendicular_tubes}(d). As suggested by~\cite{Yao:2020}, this is the region where one observes, at much higher Reynolds numbers, a proliferation of very small-scale motion, leading to a $k^{-5/3}$ energy spectrum. A faint signature of this can be seen in the relatively large value of $T_{4,5}$, larger than $T_{3,4}$, in the final panel of Fig.~\ref{fig:perpendicular_tubes}(a). This suggests a strong energy transfer toward bands of higher wavenumbers, which is consistent with the development of a $k^{-5/3}$ energy spectrum at large wavenumbers, observed in higher Reynolds-number reconnections by~\cite{Yao:2020}. These late-stage dynamics occur according to a cascade of instabilities \citep{Yao:2020}, possibly similar to the one documented by~\cite{McKeown:2020}.  

We also note that the strain, shown in panels (e) is most intense in the spaces between interacting vortices, even if little more can be said about it. Indeed, the correlations for this flow between $\mathcal{T}_{K,Q}$ and the filtered strains and vorticities are much weaker than for the originally anti-parallel tubes, detailed in Section \ref{sec:t-dependent-Fourier}. They are practically zero for the signed values of $\mathcal{T}_{K,Q}$ and in the range of $0.4-0.5$ for $|\mathcal{T}_{K,Q}|$ in the shells $I_2$, $I_3$, and $I_4$. This can be explained by the fact that the relationship between energy transfer and both strain and vorticity is less direct in this flow, especially as there are large regions of the flow where no appreciable energy transfer occurs. Similarly, there is a larger correlation ($0.5-0.7$) between $S_Q$ and $\omega_Q$ for $I_2$, $I_3$, and $I_4$ than what is observed previously, even though not for $I_1$. This indicates again that the first shell is rather inactive in this flow, as most of the dynamics is concentrated in a very small region.

The results of this section set a limit on how much further analysis can be done to relate the local dynamics in this system to those of the general HIT case. Although the flattening of the two interacting tubes into vortex sheets, observed in many simulations \citep{Yao:2022}, does lead to the formation of small-scale flow structures, its signature on the transfer of energy toward small scales is qualitatively different from what is seen in HIT, as detailed in Section~\ref{sec:HIT}. The faint cascade observed at later times is restricted to very small scales (i.e. large wavenumbers), where viscous effects are already significant. The transfer of energy is therefore much reduced, compared to the other flows considered, with a clear inertial regime.

\end{document}